\documentclass{aa}  

\usepackage{graphicx}
\usepackage{txfonts}
\usepackage{hyperref}
\usepackage{longtable}
\usepackage{supertabular}
\usepackage{amsmath}
\usepackage{placeins}
\usepackage{booktabs}
\usepackage{lscape}
\begin{document} 

\title{Star-spot activity, orbital obliquity, transmission spectrum, physical properties, and TTVs of the HATS-2 planetary system}
\titlerunning{The HATS-2 planetary system}
\authorrunning{F. Biagiotti et al.}
   
\author{
F. Biagiotti\inst{1,2}
\and 
L. Mancini\inst{3,4,5}
\and 
J. Southworth\inst{6}
\and 
J. Tregloan-Reed\inst{7}
\and
L. Naponiello\inst{3,8,4}
\and 
U.~G. J\o{}rgensen\inst{9}
\and 
N. Bach-M\o{}ller\inst{9}
\and
%A.~J. Barker
%\and 
M. Basilicata\inst{3,4}
\and
M. Bonavita\inst{10}
\and
V. Bozza\inst{11,12}
\and
M.~J. Burgdorf\inst{13}
\and
M. Dominik\inst{14}
\and
R. Figuera Jaimes\inst{15,16,14}
\and
Th. Henning\inst{5}
\and
T.~C. Hinse\inst{17}
\and
%J.~A. Hitchcock
%\and
M. Hundertmark\inst{18}
\and
%Y. Jongen
%\and
E. Khalouei\inst{19}
\and
%H. Korhonen
%\and
P. Longa-Pe\~{n}a\inst{20}
\and
N. Peixinho\inst{21}
\and
M. Rabus\inst{21}
\and
S. Rahvar\inst{19}
\and
S. Sajadian\inst{22}
\and
J. Skottfelt\inst{23}
\and
C. Snodgrass\inst{10}
\and
Y. Jongen\inst{24}
\and
J.-P. Vignes\inst{24}
}
\institute{
%1
Department of Physics, University of Rome ``La Sapienza'', Piazzale Aldo Moro 2, 00185 Rome, Italy \\
\email{francesco.biagiotti@uniroma1.it}
\and
%2
INAF -- Istituto di Astrofisica e Planetologia Spaziali (INAF-IAPS), Via Fosso del Cavaliere 100, I-00133, Rome, Italy
\and
% 3
Department of Physics, University of Rome ``Tor Vergata'', Via della Ricerca Scientifica 1, 00133, Rome, Italy
\and
% 4
INAF -- Turin Astrophysical Observatory, via Osservatorio 20, 10025, Pino Torinese, Italy
\and
% 5
Max Planck Institute for Astronomy, K\"{o}nigstuhl 17, 69117 -- Heidelberg, Germany
\and
% 6
Astrophysics Group, Keele University, Keele ST5 5BG, UK
\and
% 7
Instituto de Astronomia y Ciencias Planetarias de Atacama, Universidad de Atacama, Copayapu 485, Copiapo, Chile
\and
% 8
Department of Physics and Astronomy, University of Florence, Largo Enrico Fermi 5, 50125 Firenze, Italy
\and
% 9
Centre for ExoLife Sciences, Niels Bohr Institute, University of Copenhagen, \O{}ster Voldgade 5, 1350 Copenhagen,
Denmark
\and
% 10
Institute for Astronomy, University of Edinburgh, Royal Observatory, Edinburgh EH9 3HJ, UK
\and
% 11
Department of Physics ``E.R. Caianiello'', University of Salerno, Via Giovanni Paolo II 132, 84084, Fisciano, Italy
\and
% 12
Istituto Nazionale di Fisica Nucleare, Sezione di Napoli, Napoli, Italy
\and
% 13
Universit\"{a}t Hamburg, Department of Earth Sciences, Meteorological Institute, Bundesstrasse 55, 20146 Hamburg, Germany
\and
% 14
Centre for Exoplanet Science, SUPA, School of Phys. \& Astron., Univ. of St Andrews, North Haugh, St Andrews KY16 9SS, UK
\and
% 15
Millennium Institute of Astrophysics MAS, Nuncio Monsenor Sotero Sanz 100, Of. 104, Providencia, Santiago, Chile
% 16
\and
Instituto de Astrof\'isica, Pontificia Universidad Cat\'olica de Chile, Av. Vicu\~na Mackenna 4860, 7820436 Macul, Santiago, Chile
% 17
\and
University of Southern Denmark, Department of Physics, Chemistry and Pharmacy, Campusvej 55, 5230 Odense M, Denmark
\and
% 18
Astronomisches Rechen-Institut, Zentrum f\"{u}r Astronomie der Universit\"{a}t Heidelberg (ZAH), 69120 Heidelberg,
Germany
\and
% 19
Astronomy Research Center, Research Institute of Basic Sciences, Seoul National University, 1 Gwanak-ro, Gwanak-gu, Seoul 08826, Korea
\and
% 20
Centro de Astronom\'{i}a, Universidad de Antofagasta, Av. Angamos 601, Antofagasta, Chile
\and
% 21
Departamento de Matem{\'a}tica y F{\'i}sica Aplicadas, Facultad de Ingenier{\'i}a, Universidad Cat{\'o}lica de la Sant{\'i}sima Concepci{\'o}n, Alonso de Rivera 2850, Concepci{\'o}n, Chile
\and
% 22
Department of Physics, Isfahan University of Technology, Isfahan 84156-83111, Iran
\and
% 23
Centre for Electronic Imaging, Department of Physical Sciences, The Open University, Milton Keynes, MK7 6AA, UK
\and
% 24
ExoClock Project
}

   \date{Received ; accepted }

% \abstract{}{}{}{}{} 
% 5 {} token are mandatory
 
\abstract
  % context heading (optional)
  % {} leave it empty if necessary  
{}
%{Precise multi-colour photometric follow-up observations are very useful to characterize poorly studied transiting-planetary systems. }
  % aims heading (mandatory)
{Our aim in this paper is to refine the orbital and physical parameters of the HATS-2 planetary system and study transit timing variations and atmospheric composition thanks to transit observations that span more than ten years and that were collected using different instruments and pass-band filters. We also investigate the orbital alignment of the system by studying the anomalies in the transit light curves induced by starspots on the photosphere of the parent star.}
  % methods heading (mandatory)
{We analysed new transit events from both ground-based telescopes and NASA's TESS mission. 
%19 new transit events of HATS-2\,b with the Danish 1.54\,m telescope and with the GROND multi-band camera at the MPG\,2.2\,m telescope. We also analysed published and new data sets from the ExoClock archive and the photometric data collected by NASA’s TESS space telescope over three Sectors during its primary and extended missions.
Anomalies were detected in most of the light curves and modelled as starspots occulted by the planet during transit events. We fitted the clean and symmetric light curves with the JKTEBOP code and those affected by anomalies with the PRISM+GEMC codes to simultaneously model the photometric parameters of the transits and the position, size, and contrast of each starspot.
%, which was modelled as a single and round spot.
}
% results heading (mandatory)
{We found consistency between the values we found for the physical and orbital parameters and those from the discovery paper and  ATLAS9 stellar atmospherical models. We identified different sets of consecutive starspot-crossing events that temporally occurred in less than five days. Under the hypothesis that we are dealing with the same starspots, occulted twice by the planet during two consecutive transits, we estimated the rotational period of the parent star and, in turn the projected and the true orbital obliquity of the planet. We find that the system is well aligned.
%$\lambda=2^{\circ}.33\pm5^{\circ}.29$ and $\psi=47^{\circ}.67\pm20^{\circ}.21$, respectively.
We identified the possible presence of transit timing variations in the system, which can be caused by tidal orbital decay, and we derived a low-resolution transmission spectrum.
%We derived a limit for the modified stellar tidal quality factor of $Q^{'}_{*}>(1.99 \pm 0.26)\times 10^4$, considering the 95 per cent confidence lower limits on the measured $\dot{P}$.
%We derived a low-resolution transmission spectrum from which we measured 
%a significant variation of the planetary radius between the $g^{\prime}$ 
%and $i^{\prime}$ bands and between the $z^{\prime}$ and $i^{\prime}$ bands. 
%In both cases, the variation is roughly 8 pressure scale heights, 
%with a confidence level of more than $7\,\sigma$, and could be due to 
%absorber species (either alkali metals or oxide compounds) 
%in the atmosphere of HATS-2\,b.
}
{}
  

\maketitle
%
%------------------------------
\section{Introduction}
\label{Sec:introduction}
Due to their intrinsic size and proximity to their parent stars and despite their rarity, transiting hot Jupiters are a class of exoplanets that have received a great deal of attention from the scientific community since they were discovered \citep[see e.g.][for a comprehensive review]{2022ASSL..466....3R}. Such peculiar characteristics offer many advantages from an observational point of view, and guarantee the possibility of characterising most of the physical and orbital parameters of these planets with extreme precision. They are also the best targets for atmospheric-characterisation observations, as recently demonstrated by the JWST \citep{2023Natur.614..653A,2023Natur.614..659R,2023Natur.614..664A,2023Natur.614..670F}. 

Even so, there are still some aspects of this population that are not yet well understood and are under investigation, such as those regarding the physical mechanisms that regulate the formation, accretion, and evolution processes or those that cause their migration from the snowline ($\sim3$\,au) up to roughly $0.01$\,au from their host stars.
For example, several mechanisms have been advocated that are able to shrink the orbit of a giant planet, such as dynamical interactions through planet-planet scattering \citep{1996Sci...274..954R,2014prpl.conf..787D}, the Kozai mechanism 
\citep[e.g. ][]{2003ApJ...589..605W}, and the disc-planet interaction \citep{1996Natur.380..606L,1997ApJ...482L.211W}.
However, whatever the responsible mechanism is, the planetary orbital eccentricity, $e$, and the spin-orbit obliquity, $\psi$, should be affected. If planet-planet scattering is the main mechanism that produces hot Jupiters, these scattering encounters should randomise the orbital planes, whereas disc-planet interactions force the planet's orbit to be
coplanar throughout its migration journey and we should observe an excess of flat architectures
\citep[see e.g.][]{2022ASSL..466..143B}.

In this context, we have been carrying out a more-than-ten-year project \citep{2009MNRAS.396.1023S} to better characterise this class of planets and with the aim of measuring their physical and orbital parameters with an accuracy of less than $5\%$.
Our project is mainly based on high-quality photometric observations of planetary-transit events and, in several cases, was supported by high-resolution spectroscopic observations to characterise the atmospheric properties and activity host star \citep[see e.g.][]{2014A&A...562A.126M} or to measure the Rossitter-McLaughlin effect, which allows the measurement of the project spin-orbit alignment, $\lambda$, \citep[see e.g.][]{2018A&A...613A..41M,2022A&A...664A.162M}.

In a few cases we faced the presence of anomalies on the transit light curves in the form of small bumps, which are attributable to the presence of single starspots or groups of starspots on the photosphere of the parent stars \citep[see e.g.][]{2015A&A...579A.136M,2017MNRAS.465..843M}. These bumps appear when the planet, during its transits, temporarily hides the starspots, thus obscuring colder regions of the photosphere \citep{2003ApJ...585L.147S,2010exop.book...55W}. Appropriate modelling of the light curve allows the parameters of the starspot, to be estimated, under the simplistic hypothesis that this is a single circular spot.

In the case that the orbit of the transiting planet is well aligned with the spin of its parent star and the planet has an orbital period much shorter than the stellar rotation period, then the planet can occult the same starspot multiple times.
%the planet has an orbital period of a few days, 
%while its parent star has a rotation period of roughly a %month, the planet can occult two or three times the same %starspot. 
In these circumstances, the measurement of $\lambda$ becomes feasible with photometric monitoring of two consecutive transits \citep{2013MNRAS.428.3671T}. Instead, if there is a spin-orbit misalignment, the planet will not occult the same starspot again at the subsequent transit, and we can at most obtain a lower limit on the stellar obliquity \citep{2017AJ....153..205D}.

In this work we report the results of our study of the planetary system HATS-2 that hosts a transiting hot Jupiter, HATS-2\,b
($M_{\rm p}\approx 1.3 \, M_{\rm Jup}$; $R_{\rm p}\approx 1.2 \, R_{\rm Jup}$), which orbits around a K\,V dwarf star ($V=13.6$\,mag) in roughly 1.35\,days \citep{2013A&A...558A..55M}. The authors of this discovery reported multi-band light curves of two transits in which this planet occulted starspots and, for the first time ever as far as we know, even a bright chromospheric region known as stellar {\it plage}. Using two ground-based telescopes at the ESO Observatory of La Silla, we obtained new high-quality light curves of the HATS-2\,b transits and, thanks to TESS photometric data, we recognised new starspots and characterised their properties. We also reviewed the main physical and orbital parameters of this planetary system and probed the chemical composition of the planetary atmosphere. Possible transit time variations (TTVs) were, also investigated, using these new and archival data.

The paper is organised as follows. In Sect.~\ref{sec:2} we present the new photometric follow-up observations used to characterise the system and the relative data-reduction procedure. In Sect.~\ref{sec:light_curve_analysis} we present the analysis of the light curves, while in Sect.~\ref{sec:prop} we revise the main physical properties of the HATS-2 planetary system, focusing on the analysis of the observed starspots events. We analyse the transit times variations in Sect.~\ref{sec:ttv}, while in Sect.~\ref{sec:atm1} we investigate the variation in the planetary radius as a function of wavelength. In Sect. \ref{sec:atm2} we analyse the TESS phase curves. Finally, we summarise our results in Sect. \ref{sec:sum}.

%------------------------------

%-----------------------------------
\section{Observations and data Reduction}
\label{sec:2}
%-----------------------------------
\begin{figure*}
    \centering
   %\resizebox{\hsize}{!}{\includegraphics{multidanish.eps}}
    \includegraphics[width=14cm]{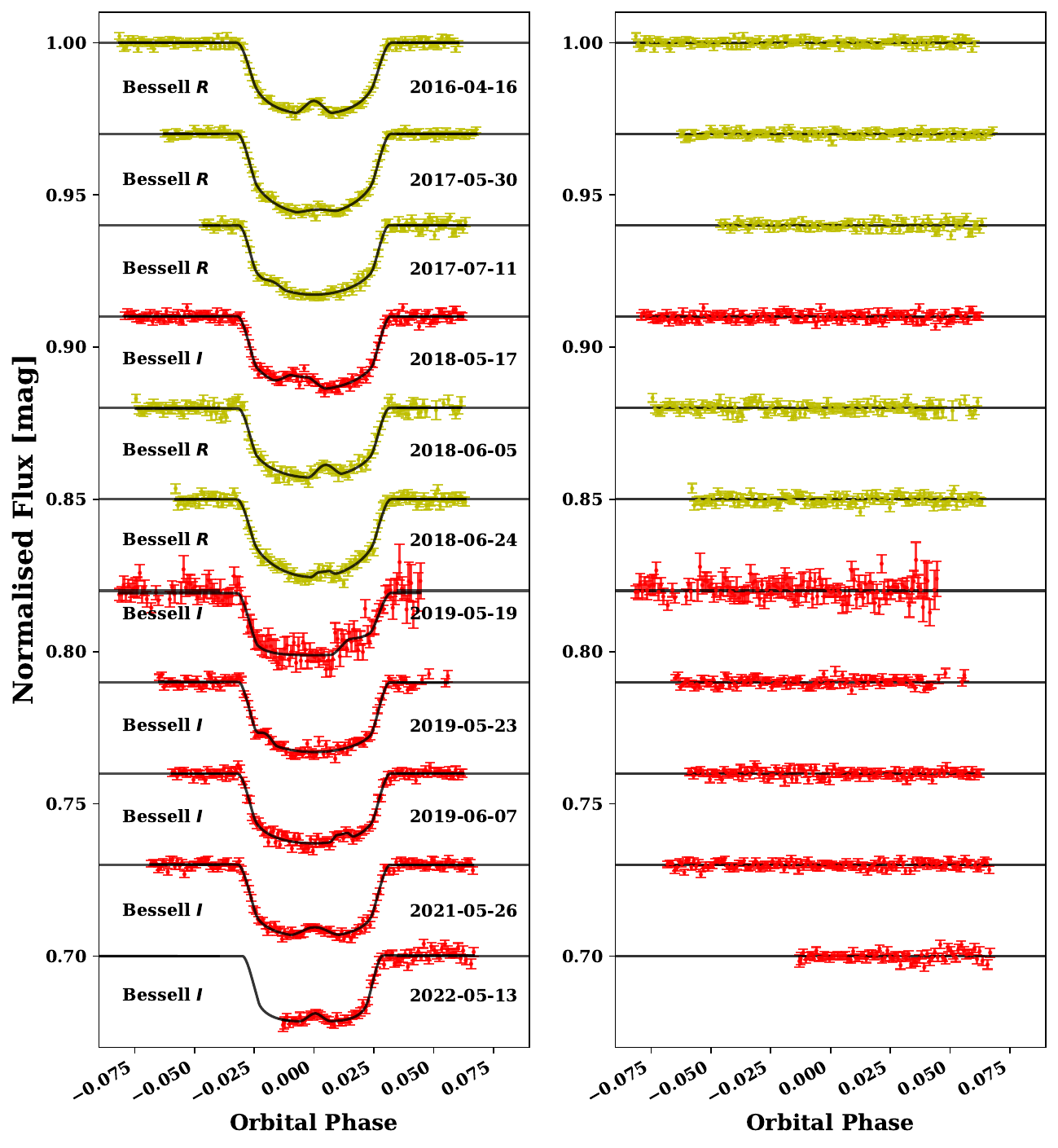}
\caption{Transits from the DK telescope. {\it Left panel}: Light curves of 11 transits (ten complete and one partial) of HATS-2\,b observed with the Danish 1.54\,m telescope, shown in chronological order. They are plotted against the orbital phase and are compared to the best-fitting models. Labels indicate the observation date and the filter (Bessell $I$ and $R$) that was used for each data set. Starspot anomalies are visible in all the light curves.
{\it Right-hand panel}: The residuals of each fit.}
\label{fig:dan}
\end{figure*}
\begin{figure*}
    \centering
    \vspace{-0.5cm}
    \includegraphics[width=14cm]{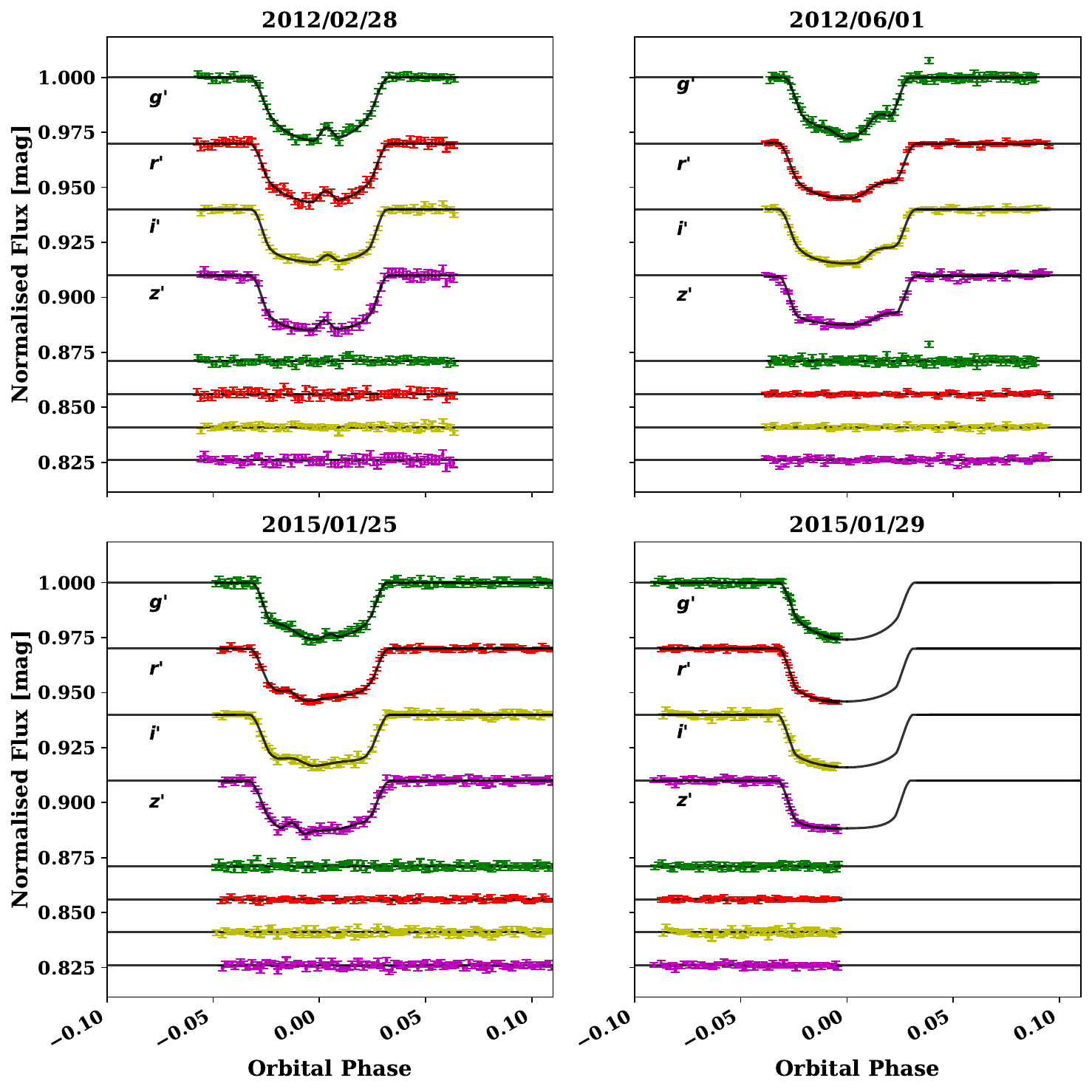}
\caption{Light curves of four (three complete and one partial) transits of HATS-2\,b simultaneously observed in four optical bands (Sloan $g^{\prime}$, $r^{\prime}$, $i^{\prime}$, $z^{\prime}$) with the GROND multi-band camera at the MPG\,2.2\,m telescope. They are shown in date order. The light curves in the top panels are from \citet{2013A&A...558A..55M}, while those in the bottom panels are from this work. Starspot anomalies are visible in the first three data sets. The light curves are plotted against the orbital phase and are compared to the best-fitting models. The residuals of the fits are shown at the base of each panel.}
\label{fig:grond}
\end{figure*}
This work is mainly based on photometric follow-up observations of transit events of the exoplanet HATS-2\,b, which were performed using ground-based and space telescopes. The ground-based data were collected starting in 2015, using the DFOSC camera, installed at the Danish 1.54\,m telescope and the GROND multi-band camera, which is one of the instruments at the MPG\,2.2\,m telescope and provides data in different passband filters simultaneously. The observations were, therefore, carried out through different optical passbands (covering the $400-1000$\,nm wavelength range) to also investigate possible variations in the transit depth with the wavelength and the dependence of the starspot contrast with wavelength. 

We observed 11 new transits with the Danish 1.54\,m telescope and 2 new transit events with the MPG\,2-2\,m telescope for a total of 19 new optical light curves. All the observed transits were fully monitored except for the transit on 2022 May 26 observed with the Danish telescope and the transit on 2015 January 29 observed with the MPG\,2.2\,m telescope. A summary of the observations with their properties and observing conditions (airmass through the night and moon illumination) is given in Table~\ref{tab:obs}.
We also reanalysed the two transit light curves observed with GROND, which were reported in the discovery paper \citep{2013A&A...558A..55M}.

Furthermore, HATS-2 was monitored by the \emph{TESS} space telescope \citep{2014SPIE.9143E..20R} with the 2 min cadence during Sector 10 of its primary mission and during Sectors 36 and 63 of its extended mission. 

Many of the transits recorded by \emph{TESS} show anomalies that most likely are connected to occultations of starspots by the planet. Similar anomalies are also present in all the complete transit light curves recorded with the Danish and MPG\,2.2\,m telescopes.  
\begin{table*}
\centering
\caption{Details of the ground-based transit observations presented in this work.  }
\label{tab:obs}
\resizebox{\hsize}{!}{  
\begin{tabular}{c c c c c c c c r c c c}
\hline \hline \\ [-8pt]
Date of  & Instrument & Start and & $N_{\rm obs}$ & $T_{\rm exp}$ & $T_{\rm obs}$ & Filter & Airmass & Moon & Aperture & $N_{\rm poly}$ & Scatter\\
first obs. & & end times (UT) & & (s) & (s) & & & illum.& sizes (px) &  & (mmag) \\
\hline \\ [-6pt]
% 
%2012/02/28 & GROND & & 72 & & & $g'$ & & & & &1.06\\
%2012/02/28 & GROND && 72 & &  &$r'$ & &  & & &1.22\\
%2012/02/28 & GROND && 72 & &  &$i'$ & &  & & &0.95\\
%2012/02/28 & GROND && 72 & &  &$z'$ & &  & & &1.24\\
%2012/06/01 & GROND && 100 & &  &$g'$ & &  & & &1.24\\
%2012/06/01 & GROND && 100 & &  &$r'$ & &  & & &0.73\\
%2012/06/01 & GROND && 100 & &  &$i'$ & &  & & &0.92\\
%2012/06/01 & GROND && 100 & &  &$z'$ & &  & & &1.17\\
2015/01/26 & GROND & 03:52  $\rightarrow$ 08:59 & 117 & 90 & 157 &$g'$ & 1.90$\rightarrow$1.00$\rightarrow$ 1.02 & 40\% & 23,80,100 & 1 & 1.07\\
2015/01/26 & GROND & 03:52  $\rightarrow$ 08:59 & 117 & 90 & 157 &$r'$ & 1.90$\rightarrow$1.00$\rightarrow$ 1.02 & 40\% & 25,80,100 & 1 & 0.73\\
2015/01/26 & GROND & 03:52  $\rightarrow$ 08:59 & 117 & 90 & 157 &$i'$ & 1.90$\rightarrow$1.00$\rightarrow$ 1.02 & 40\% & 23,80,100 & 1 & 1.03\\
2015/01/26 & GROND & 03:52  $\rightarrow$ 08:59 & 117 & 90 & 157 &$z'$ & 1.90$\rightarrow$1.00$\rightarrow$ 1.02 & 40\% & 23,80,100 & 1 & 1.12\\
2015/01/30 & GROND & 03:54 $\rightarrow$ 06:42 & 76 & 85 & 106 &$g'$ & 1.60$\rightarrow$1.05 & 80\% & 20,80,100 & 1 & 0.88\\
2015/01/30 & GROND & 03:54 $\rightarrow$ 06:42 & 73 & 85 & 106 &$r'$ & 1.60$\rightarrow$1.05 & 80\% & 23,80,100 & 1 & 0.60\\
2015/01/30 & GROND & 03:54 $\rightarrow$ 06:42 & 72 & 85 & 106 &$i'$ & 1.60$\rightarrow$1.05 & 80\% & 23,80,100 & 1 & 1.01\\
2015/01/30 & GROND & 03:54 $\rightarrow$ 06:42 & 75 & 85 & 106 &$z'$ & 1.60$\rightarrow$1.05 & 80\% & 23,80,100 & 1 & 0.81\\
2016/04/16 & DFOSC & 23:33 $\rightarrow$ 03:59 & 149 & 100 & 112 & $R$ & 1.08$\rightarrow$1.05$\rightarrow$1.32 & 77$\%$ & 15,30,50 & 1 & 0.96 \\
2017/05/30 & DFOSC & 22:49 $\rightarrow$ 03:03 & 137 & 100 & 113 & $R$ & 1.02$\rightarrow$1.02$\rightarrow$2.47 & 34$\%$ & 15,33,50 & 2 & 0.88 \\
2017/07/11 & DFOSC &22:54 $\rightarrow$ 02:28 & 103 & 100 & 125 & $R$ & 1.49$\rightarrow$1.63 & 92$\%$ & 17,35,55 & 2 & 1.21 \\
2018/05/17 & DFOSC &00:02 $\rightarrow$ 04:38 & 109 & 100 & 112 & $I$ & 1.02$\rightarrow$1.08$\rightarrow$1.63 & 9$\%$ & ~~~9,17,40 & 2 & 1.13 \\
2018/06/05 &DFOSC & 23:15 $\rightarrow$ 03:38 & 123 & 100 & 122 & $R$ & 1.07$\rightarrow$1.10$\rightarrow$4.94 & 57$\%$ & 15,33,53 & 2 & 1.38 \\
2018/06/24 & DFOSC &22:47 $\rightarrow$ 02:43 & 127 & 100 & 112 & $R$ & 1.17$\rightarrow$1.22 & 90$\%$ & 17,35,57 & 3 & 1.29 \\
2019/05/19 & DFOSC &23:16 $\rightarrow$ 03:59 & 122 & 100 & 111 & $I$ & 1.01$\rightarrow$1.02$\rightarrow$2.72 & 98$\%$ & 10,24,40 & 4 & 2.80 \\
2019/05/23 & DFOSC &01:18 $\rightarrow$ 05:19 & 112 & 100 & 115 & $I$ & 1.25$\rightarrow$1.32 & 74$\%$ & 14,32,45 & 2 & 1.23 \\
2019/06/07 & DFOSC &23:00 $\rightarrow$ 02:57 & 121 & 100 & 116 & $I$ & 1.06$\rightarrow$1.09$\rightarrow$3.10 & 25$\%$ & 13,33,55 & 3 & 1.12 \\
2021/05/26 & DFOSC &23:46 $\rightarrow$ 04:09 & 134 & 100 & 116 & $I$ & 1.06$\rightarrow$1.09$\rightarrow$4.35 & 99$\%$ & 13,34,52 & 3 & 1.03 \\
2022/05/23 & DFOSC &03:00 $\rightarrow$ 06:00 & ~~73 & 100 & 129 & $I$ & 1.01$\rightarrow$1.57$\rightarrow$2.00 & 94$\%$ & 18,40,60 & 2 & 1.42 \\
\hline
\multicolumn{11}{c}{\footnotesize Notes: $N_{\rm obs}$ is the number of observations, $T_{\rm exp}$ is the exposure time, $T_{\rm obs}$ is the observational cadence, and `Moon illum.' is the geocentric fractional}\\
\multicolumn{11}{c}{\footnotesize  illumination of the Moon at midnight. The aperture sizes are the radii of the software apertures for the star, inner sky and outer sky, respectively.}\\
\multicolumn{11}{c}{\footnotesize  Scatter is the rms scatter of the data versus a fitted model.}\\
\end{tabular}
}
\end{table*}

%-----------------------------------
\subsection{Danish 1.54 m telescope}
\label{sec:DK154}
%-----------------------------------  
Five transits of HATS-2\,b were observed through a Bessell-$R$ filter, while six transits (five complete and one incomplete) were observed through a Bessell-$I$ filter with the DFOSC (Danish Faint Object Spectrograph and Camera) imager mounted on the Danish 1.54\,m Telescope at the ESO Observatory in La Silla (Chile). The transit recorded on 2022 May 26 was not fully covered due to technical problems at the beginning of the observations, whereas the observations performed on 2019 May 19 were severely affected by weather conditions.

The DFOSC is equipped with a focal reducer, which changes the original focal length ($\approx$13\,m) of the telescope and allows for a wider field of view on the sky. The current camera of the DFOSC is an e2v CCD with $2048 \times 4096$ pixels, a plate scale of $0.39^{\prime \prime}$ per pixel and 32-bit encoding. 
%It has a gain of $0.25 \,e^{-}/$\,ADU, a RON of %$4.5 \,e^{-}$, and it holds a linear response up %to $\simeq 60\,000$\,ADU. 
In its current set-up, only half of the CCD is illuminated by the incoming starlight, so the actual FOV is $13.7^{\prime}\times13.7^{\prime}$. The CCD has been further windowed in many of the observations to reduce the readout time and, therefore, improve the sampling. 

The data were analysed by using the IDL/ DEFOT pipeline \citep{2009MNRAS.396.1023S,2012MNRAS.426.1338S}. Following a standard approach, which was already adopted in the previous works of the series \citep[e.g. see][]{2012MNRAS.420.2580S,2017AJ....153..191S}, we first calibrated the raw science images by master-bias and master-flat frames; these frames were obtained by median combining a set of individual bias and sky flat-field images, which were taken on the same night as each transit observation. Then, we performed aperture photometry selecting an ensemble of comparison stars and cross-correlating each image in a time series against a reference image. In this step, the time stamps were converted from JD$_{\rm UTC}$ to BJD$_{\rm TDB}$, following the convention suggested by \citet{2010PASP..122..935E}. Once the aperture photometry analysis was completed, differential-magnitude light curves were generated against each comparison star for each time series. Each curve has a dimension equal to the number of observations $N_{\rm obs}$. For the purpose of detrending the data, we fitted a polynomial with order between $0<N_{\rm poly}<5$ to all the points before the ingress and after the egress, for the set of the $n$ light curves. The parameters of the fit were the $n\cdot (N_{\rm poly}+1)$ coefficients and the $n$ weights, while the number of data points fitted are $n\cdot N_{\rm obs}$. The weights were simultaneously optimised to minimise the scatter in the out-of-transit data points. Finally, all the points were combined into one ensemble by a weighted flux summation, obtaining the final light curve for each data set. All the light curves are shown in Fig.~\ref{fig:dan}. 

The aperture radii, the order of the detrending polynomial, and the rms scatter values of the data against a best-fitting model (see next section) are listed in Table~\ref{tab:obs}, showing the average good quality of these new ground-based data. In all the final light curves, see Fig.~\ref{fig:dan}, it is possible to observe the presence of higher-order photometric effects (i.e. bumps connected to starspots, as described in Sect.~\ref{sec:light_curve_analysis}).

%-----------------------------------  
\subsection{MPG 2.2\,m telescope}
%-----------------------------------  
Two transit events of HATS-2\,b were recorded on 2015 January 25 and 2015 January 28 thanks to the GROND instrument, which is mounted on the Nasmyth focus of the MPG 2.2\,m Telescope at the ESO Observatory in La Silla (Chile). GROND stands for Gamma-Ray Burst Optical/Near-Infrared Detector and is a 7-channel imager, which allows observations with different filters at the same time \citep{2008PASP..120..405G}. Because of the connection of GRBs with high-redshift galaxies, the available filters are the Sloan $g^{\prime},\,r^{\prime},\,i^{\prime},\,z^{\prime}$ bands and the near-infrared $J,\,H,\,K$ bandpass filters. The camera is essentially composed of a system of dichroics and collimators, which split the incoming light towards two arms (optical and NIR) and then in more beams so that all seven detectors are illuminated at the same time. The first division is required because of the differences between the NIR-detector arrays and optical CCDs, such as temperature sensitivity. For our work, we only used the data from the optical channel, which is composed of a system of four dichroics and four back-illuminated E2V CCDs with $2048\times2048$ pixel. For each CCD, the plate scale is $0.158^{\prime \prime}$ per pixel and the FOV is $5.4 \times 5.4$\,arcmin$^2$. 
%For the $g'$ channel the gain is 1.45 $e^- /$ADU %and the RON is 4.5 $e^-$, for the $r'$ channel %the gain is 1.33 $e^- /$ADU and the RON is 5.75 %$e^-$, for the $i'$ channel the gain is 1.62 %$e^- /$ADU and the RON is 5.35 $e^-$, and for %the $z'$ channel the gain is 1.74 $e^- /$ADU and %the RON is 7.0 $e^-$.

The first transit was completely monitored, whereas the second was only partially monitored due to a crash of the TCS (telescope control system). The observations were reduced in the same way as those from the Danish Telescope (see Sect.\ref{sec:DK154}). Other two transit events obtained with GROND were reported in the discovery paper \citep{2013A&A...558A..55M} and reanalysed here. We have not re-performed photometry for these observations. The 16 GROND light curves are shown in Fig.~\ref{fig:grond}.

%-----------------------------------  
\subsection{TESS space telescope}
\begin{figure*}
\centering
   \resizebox{\hsize}{!}{\includegraphics{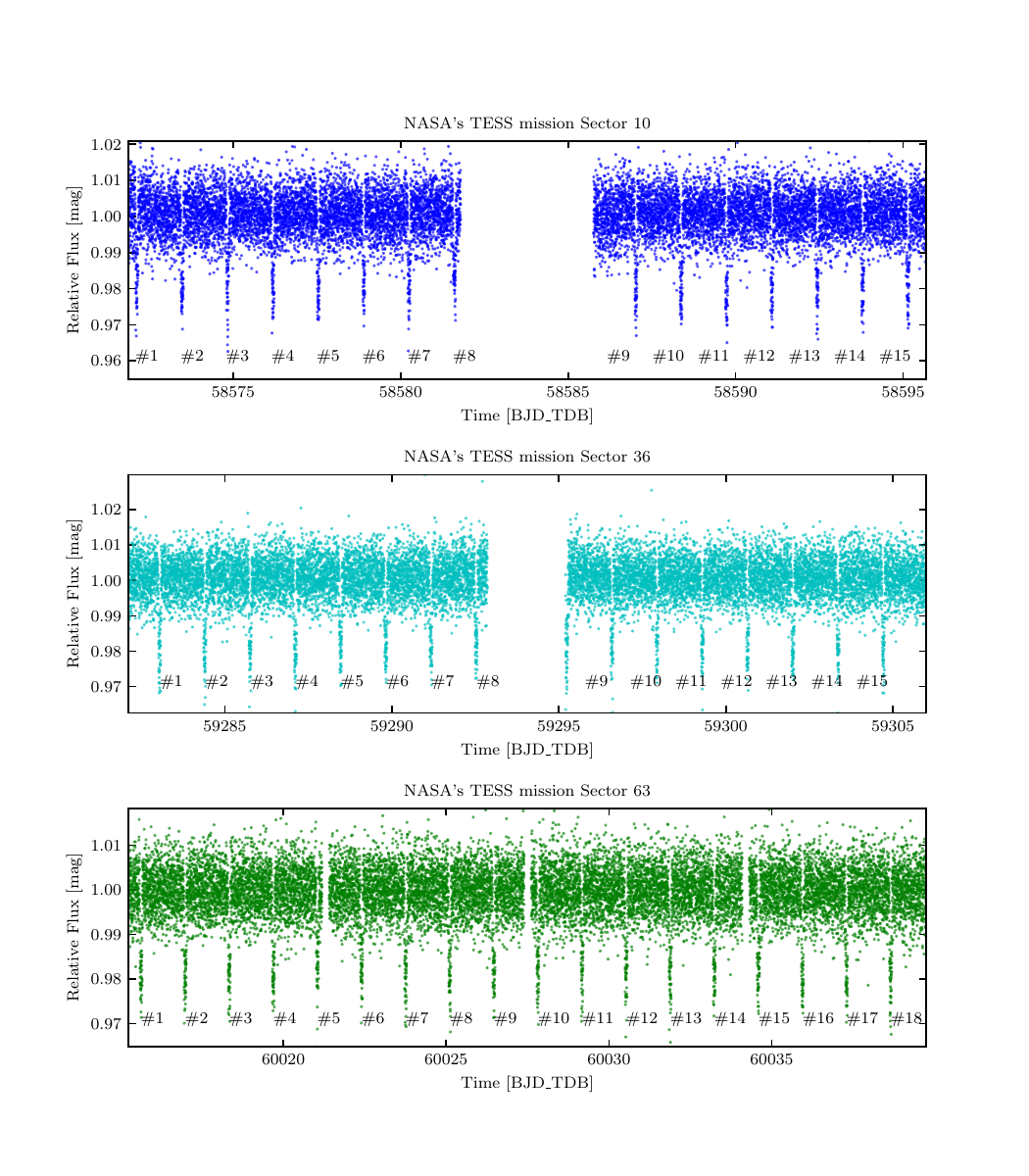}}
\caption{Photometric monitoring of HATS-2 by the TESS space telescope. {\it Top panel}: Data from Sector 10. {\it Middle panel}: Data from Sector 36. {\it Bottom panel}: Data from Sector 63. Fifteen transit events of HATS-2\,b were detected by TESS in Sector 10 and 15 in Sector 36, while 18 transit events were detected in Sector 63.}
\label{fig:tess}
\end{figure*}

%-----------------------------------  
The PDCSAP
\citep[Pre-search Data Conditioning Simple Aperture Photometry;][]{2012PASP..124.1000S,2014PASP..126..100S}
light curve of HATS-2 from both Sector 10 and 36 of the NASA \emph{TESS} mission was downloaded through the Python package \texttt{lightkurve}  \citep{2018ascl.soft12013L}, from the Mikulski Archive for Space Telescopes (MATS). Using the same Python package, we downloaded also raw PDCSAP data from Sector 63, for which the detrending was performed thanks to the {\texttt{wotan}} Python package \citep{wotan}. The detrended data from each sector are shown in Fig. \ref{fig:tess}. Continuous observations of the target star were obtained for two periods of roughly  25 days (from 2019 March 28 to 2019 April 21)  and 24 days (from 2021 March 08 to 2021 March 31) for Sectors 10 and 36, respectively. For Sector 63, the target star was observed for a period of 25 days (from 2023 March 11 to 2023 April 05) collecting, however, more transits. In fact, the data from Sector 10 and Sector 36 contain fifteen transits of HATS-2\,b, while Sector 63 contains eighteen transits. The sub-optimal quality of TESS data is due to the relative faintness of the HATS-2 star ($V = 13.6$\,mag), which is at the limit of the working magnitude of TESS. Nevertheless, several transits show clear anomalies, see Figs.~\ref{fig:tess1},~\ref{fig:tess3},~\ref{fig:tess4}.
   
%----------------------------
\section{Light curves analysis}
\label{sec:light_curve_analysis}
%----------------------------

Since many of our obtained light curves contained anomalies that could be connected to starspot complexes occulted by the exoplanet HATS-2\,b during its transits in front of its parent star, we had to model them with codes that are optimised against the presence of starspots and plages. Previous works showed that the presence of starspot anomalies not properly modelled could cause measures that mimic TTVs \citep{2013A&A...556A..19O}, or an anomalous planetary radius towards the bluest wavelengths \citep{2014A&A...568A..99O}. 

The quality of our light curves is good enough to let us hypothesise that the anomalies are caused by starspot-occultation events (this is also supported by the spectral class of the host star) but not good enough to make assumptions about the fine structures of the starspots (as discriminate between single spots or groups of spots). So, it is normal practice to model the anomalies assuming one or more circular spots. As in the previous works of our series \citep[e.g.][]{2013MNRAS.436....2M,2014MNRAS.443.2391M,2015A&A...579A.136M,2017MNRAS.465..843M}, we decided to use the PRISM and GEMC codes\footnote{https://github.com/JTregloanReed/PRISM$\_$GEMC} \citep{2013MNRAS.428.3671T,2015MNRAS.450.1760T}, which let us model the light curves fitting both the transit and one or several spot-crossing events simultaneously thanks to a bayesian approach and the use of genetic algorithms \citep{ter}. PRISM+GEMC is set to use the quadratic Limb-Darkening (LD) law and, assuming a circular orbit, can model a light curve fitting for the following parameters: the sum of the fractional radii of the star and planet\footnote{$r_{\star}$ and $r_{\rm p}$
are defined as the stellar radius $R_\star$ and the planetary radius $R_\mathrm{p}$ scaled by the semi-major axis $a$, respectively.}, $r_{\star}+r_{\rm p}$, the ratio of the radii, $k=\frac{R_{\rm p}}{R_{\star}}=\frac{r_{\rm p}}{r_{\star}}$, the orbital inclination, $i$,
%the orbital eccentricity, $e$, the longitude of the %periastron, $\omega$, 
the LD coefficients, $u_1$ and $u_2$, the time of mid-transit, $T_0$, the longitude and latitude ($\phi \,, \theta$) of the centre of the starspot and, finally, the radius, $r_{\rm s}$, and the contrast, $\rho$, of the starspot (i.e. the ratio of the surface brightness of the starspot to that of the surrounding photosphere).

We used PRISM+GEMC to analyse all the ground-based data sets presented in the previous section, four transits from Sector 10 of the TESS mission, six transits from Sector 36, and six transits from Sector 63. We also reanalysed the two light curves that were recorded with the MPG\,2.2\,m telescope by \citet{2013A&A...558A..55M}. From now on, we will refer to the TESS transits from Sectors 10 and 36 using a numerical identification between 1 and 15 for both sectors, while those from Sector 63 between 1 and 18. The transit events have been labelled following a chronological order, in other words, from the left to the right referring to Fig.~\ref{fig:tess}. All the transit events, with their relative id number, are shown in Fig.~\ref{fig:tess}. 

By using PRISM+GEMC, we analysed the light curves $\#$2, $\#$4, $\#$14, and $\#$15 from Sector 10 and those from $\#$6 to $\#$11 from Sector 36, as they show clear transit anomalies. Regarding Sector 63, the curves that presented anomalies were the ones labelled: $\#3$, $\#9$, $\#11$, $\#13$, $\#14$, and $\#15$. 
We used
%a stellar disc radius of 15, 
1000 generations and 256 chains during the burn-in stage, and 128 chains during the burn-out stage \citep[for details see][]{2015MNRAS.450.1760T}.
The uncertainties have been adjusted by multiplying each data set by the square root of the reduced chi-squared value obtained during the burn-in stage, that is $\beta=(\chi_{\rm red}^2)^{1/2}$.

We always fitted for a single starspot on the stellar disc, except for the cases of the 2012 June 1st $g'$ transit from the MPG\,2.2\,m, the 2015 January 25 $g'$ $r'$ $i'$ transits from the MPG\,2.2\,m, and the $\#9$ TESS transit from Sector 36. The results of the fit are listed in Tables~\ref{tab:fitsp} and \ref{tab:fit}, while the best-fitting models and their residuals are shown in Figs~\ref{fig:dan} and \ref{fig:grond}, and in the Appendix (see Figs.~\ref{fig:tess1},~\ref{fig:tess3},~\ref{fig:tess4},~\ref{fig:tess6},~\ref{fig:tess7}). Regarding Sector 63, unbinned and 10 min cadence binned data were available, and we decided to fit both for the curves that contained anomalies. In Fig.\ref{fig:tess6} and ~\ref{fig:tess7} the reader can observe both the unbinned and 10 min cadence binned data. The results were always consistent for the two different data sets and so we kept the ones that gave us the smaller uncertainties regarding the starspot properties. It is worth underlining that if we did not have the unbinned data as a comparison, deciding to keep results from binned data would have been dangerous as the binning process could introduce systematics and bias the results (e.g. \citealt{kipping10}).

The values in Tables~\ref{tab:fitsp} refer to the median values of the posterior-probability-density (PDF) function of the parameters and not to the best-fitting parameters. The large uncertainties in the starspot parameters are caused by the fact that in some cases (as for the 2017 May 30 transit) the anomalies amplitudes are too small ($<2$\,mmag) and in other cases (TESS) the scatter in the light curves is too high ($>2$\,mmag). In both scenarios, PRISM/GEMC can find reasonable starspot parameters but the associated error bars are large because a wide range of solutions could produce such small or unclear anomalies. 

The other light curves of Sector 10 and 36, which do not show detectable anomalies, have been analysed by using JKTEBOP \citep{2008MNRAS.386.1644S}, which is faster than PRISM+GEMC and can fit the same photometric parameters except for the starspot parameters. The results are listed in Table~\ref{tab:fit} and the reader can inspect the best-fitting models in Fig.~\ref{fig:tess2} and Fig.~\ref{fig:tess5}. For Sector 63, all the light curves (unbinned data) were analysed by using PRISM+GEMC. The results are listed in Table~\ref{tab:fit} and shown in Fig.~\ref{fig:tess8}.

Due to the large scatter in TESS data, it is convenient to fit multiple transits simultaneously, as pointed out by several authors
\citep[e.g. ][]{2016A&A...590A.112M,2022MNRAS.515.3212S}. This approach leads to more precise measurements of the sum and ratio of the radii, the inclination and, above all, the mid-transit times. In particular, we chose to perform a fit for each of the four periods of continuous observations. We considered transits from $\#1$ to $\#8$ and from $\#9$ to $\#15$ for the data in both Sectors 10 and 36, see Fig.~\ref{fig:tess}. We applied the same reasoning to the Sector 63 TESS data. In particular, we chose to perform four different fits considering in order: ($i$) transits from $\#1$ to $\#5$, ($ii$) from $\#6$ to $\#9$, ($iii$) from $\#10$ to $\#14$, and ($iv$) from $\#15$ to $\#18$.
%
%When a simultaneous fit is performed, the orbital period could not be fixed %anymore but must be fit as well. Moreover, JKTEBOP needs one of the %contained mid-transit times to be set as the reference, while the others %have been fitted using a first-order polynomial together with the orbital %period. The reference mid-transit times were chosen as follows:
%
%\begin{itemize}
%\item [($i$)] the mid-transit time of the $\#5$ transit; 
%\item [($ii$)] the mid-transit time of the $\#12$ transit;
%\item [($iii$)] the mid-transit time of the $\#4$ transit; 
%\item [($iv$)] the mid-transit time of the $\#14$ transit. 
%\end{itemize}
%
%This choice was made considering light curves that do not present prominent %anomalies. 
Moreover, we fixed the non-linear LD coefficient, while the linear one has been set as a free parameter. Uncertainties were estimated using a Monte Carlo method.
The final values of the parameters were estimated by a weighted mean of all the fits of all the light curves that are listed in { Table~(\ref{tab:fit})}.
%We considered the results for $r_{\star}$ and $r_{\rm p}$ individually, %performed a weighted mean on them and, then, obtained a final weighted %value for the sum of the scaled radii and the ratio of the radii. Regarding %JKTEBOP, these values are present in the output file contrary to the %PRISM/GEMC. In this case, we calculated $r_{\star}$ and $r_{\rm p}$, and %propagated the uncertainties using a Monte Carlo method, because standard %error propagation does not work well for these quantities. 

\begin{table*}
\centering 
\caption{Starspot parameters derived from the PRISM+GEMC fits of the transit light curves.}
\label{tab:fitsp}
\resizebox{\hsize}{!}{             
\begin{tabular}{c c c c c c c c c }
\hline \hline  \\ [-8pt]
Telescope & Date or ID & Filter & starspot & $\phi\,(^\circ)$ & $\theta\,(^\circ)$ & $r_{\rm s}\,(^\circ)$ & $\rho$ & $T_{\rm spot}$ (K) \\ 
\hline \\ [-6pt]
MPG\,2.2\,m & 2012/02/28 & $g'$ & $\# 1$ & $7.11 \pm 0.84$ &$5.24 \pm 4.64$ &$4.94 \pm 1.90$ & $0.11 \pm 0.16$ & $3792 \pm 739$\\
MPG\,2.2\,m & 2012/02/28 & $r'$ & $\# 1$ & 5.28 $\pm$ 1.51 &5.30 $\pm$ 5.72 &6.65 $\pm$ 3.63 &0.45 $\pm$ 0.23 &4432 $\pm$ 504 \\
MPG\,2.2\,m & 2012/02/28 & $i'$ & $\# 1$ & 7.65 $\pm$ 1.80 &3.12 $\pm$ 6.03 &5.00 $\pm$ 3.37 &0.43 $\pm$ 0.27 &4247 $\pm$ 659 \\
MPG\,2.2\,m & 2012/02/28 & $z'$ & $\# 1$ & 4.59 $\pm$ 4.28 &5.87 $\pm$ 10.80 &5.79 $\pm$ 3.52 &0.44 $\pm$ 0.34 &4136 $\pm$ 883 \\ [2pt]
MPG\,2.2\,m & 2012/06/01 & $g'$ & $\begin{tabular}{c}
$\# 1$ \\
$\# 2$ \\
\end{tabular}$ &
$\begin{tabular}{c}
36.15 $\pm$ 2.95 \\
2.26 $\pm$ 4.33 \\
\end{tabular}$
&
$\begin{tabular}{c}
4.59 $\pm$ 5.25 \\
8.48 $\pm$ 6.17 \\
\end{tabular}$
&  
$\begin{tabular}{c}
17.14 $\pm$ 4.45 \\
15.24 $\pm$ 6.74 \\
\end{tabular}$
&  
$\begin{tabular}{c}
0.64 $\pm$ 0.13 \\
1.21 $\pm$ 0.10 \\
\end{tabular}$
&  
$\begin{tabular}{c}
4846 $\pm$ 237 \\
5405 $\pm$ 182 \\
\end{tabular}$ \\
MPG\,2.2\,m & 2012/06/01 & $r'$ & $\# 1$ & 34.94 $\pm$ 1.43  &7.28 $\pm$ 6.64 &18.41 $\pm$ 2.72 &0.79 $\pm$ 0.04 &4964 $\pm$ 140 \\
MPG\,2.2\,m & 2012/06/01 & $i'$ & $\# 1$ & 33.95 $\pm$ 1.23  &7.38 $\pm$ 5.77 &17.57 $\pm$ 2.49 &0.84 $\pm$ 0.07 &4986 $\pm$ 194 \\
MPG\,2.2\,m & 2012/06/01 & $z'$ & $\# 1$ & 39.95 $\pm$ 2.11 &32.36 $\pm$ 8.41 & 21.48 $\pm$ 5.35 & 0.81 $\pm$ 0.06 &4870 $\pm$ 180 \\  [2pt]
MPG\,2.2\,m & 2015/01/25 & $g'$ 
&  
$\begin{tabular}{c}
$\# 1$ \\
$\# 2$ \\
\end{tabular}$
&  
$\begin{tabular}{c}
-36.95 $\pm$ 4.52 \\
9.27 $\pm$ 31.57 \\
\end{tabular}$
&  
$\begin{tabular}{c}
52.95 $\pm$ 13.98 \\
9.60 $\pm$ 7.82 \\
\end{tabular}$
&  
$\begin{tabular}{c}
36.63 $\pm$ 12.44 \\
2.92 $\pm$ 3.66 \\
\end{tabular}$
&  
$\begin{tabular}{c}
0.39 $\pm$ 0.23 \\
0.43 $\pm$ 0.28 \\
\end{tabular}$
&  
$\begin{tabular}{c}
4492 $\pm$ 465 \\
4559 $\pm$ 522 \\
\end{tabular}$ \\
MPG\,2.2\,m & 2015/01/25 & $r'$
&  
$\begin{tabular}{c}
$\# 1$ \\
$\# 2$ \\
\end{tabular}$
&  
$\begin{tabular}{c}
-30.74 $\pm$ 2.77 \\
18.16 $\pm$ 7.29 \\
\end{tabular}$
&  
$\begin{tabular}{c}
6.85 $\pm$ 5.40 \\
31.29 $\pm$ 17.26 \\
\end{tabular}$
&  
$\begin{tabular}{c}
10.12 $\pm$ 5.39 \\
25.37 $\pm$ 13.51 \\
\end{tabular}$
&  
$\begin{tabular}{c}
0.80 $\pm$ 0.09 \\
0.90 $\pm$ 0.09 \\
\end{tabular}$
&  
$\begin{tabular}{c}
4977 $\pm$ 206 \\
5106 $\pm$ 202 \\
\end{tabular}$ \\
MPG\,2.2\,m & 2015/01/25 & $i'$
&  
$\begin{tabular}{c}
$\# 1$ \\
$\# 2$ \\
\end{tabular}$
&  
$\begin{tabular}{c}
-27.49 $\pm$ 9.23 \\
22.13 $\pm$ 45.60 \\
\end{tabular}$
&  
$\begin{tabular}{c}
40.69 $\pm$ 15.86 \\
47.33 $\pm$ 32.72 \\
\end{tabular}$
&  
$\begin{tabular}{c}
15.71 $\pm$ 11.84 \\
21.64 $\pm$ 18.85 \\
\end{tabular}$
&  
$\begin{tabular}{c}
0.76 $\pm$ 0.18 \\
0.83 $\pm$ 0.24 \\
\end{tabular}$
&  
$\begin{tabular}{c}
4862 $\pm$ 376 \\
4974 $\pm$ 460 \\
\end{tabular}$ \\
MPG\,2.2\,m & 2015/01/25 & $z'$ & $\# 1$ & -27.29 $\pm$  1.44 &15.26 $\pm$ 7.68 &5.19 $\pm$ 3.79 & 0.62 $\pm$ 0.16 & 4541 $\pm$ 397 \\  [2pt]
DK 1.54\,m & 2016/04/16 & $R$ & $\# 1$ &-0.01 $\pm$ 0.93 &9.23 $\pm$ 6.84 &9.19 $\pm$ 3.28 & 0.69 $\pm$ 0.17 &4816 $\pm$ 328 \\
DK 1.54\,m & 2017/05/30 & $R$ & $\# 1$ &2.48 $\pm$ 5.73  &26.44 $\pm$ 16.79 &14.21 $\pm$ 11.69 &0.81 $\pm$ 0.18 &4987 $\pm$ 324\\
DK 1.54\,m & 2017/07/11 & $R$ & $\# 1$ &-37.28 $\pm$ 16.37 & 27.85 $\pm$ 20.33 &17.92 $\pm$ 17.65 &0.37 $\pm$ 0.26 & 4240 $\pm$ 622 \\
DK 1.54\,m & 2018/05/17 & $I$ & $\# 1$ &-13.91 $\pm$ 8.41 &11.72 $\pm$ 11.32 &16.22 $\pm$ 8.93  &0.84 $\pm$ 0.03 & 4977 $\pm$ 141 \\
DK 1.54\,m & 2018/06/05 & $R$ & $\# 1$ &9.80 $\pm$ 1.91 & 8.48 $\pm$ 7.60 & 9.47 $\pm$ 3.86 & 0.75 $\pm$ 0.20 & 4898 $\pm$ 368\\
DK 1.54\,m & 2018/06/24 & $R$ & $\# 1$ &8.01 $\pm$ 8.48 &6.91 $\pm$ 8.38 &4.00 $\pm$ 3.83 &0.61 $\pm$ 0.27 & 4683 $\pm$ 504\\
DK 1.54\,m & 2019/05/19 & $I$ & $\# 1$ &63.36 $\pm$ 17.20 &18.72 $\pm$ 24.66 & 42.58 $\pm$ 16.62 &0.79 $\pm$ 0.18 & 4898 $\pm$ 382\\
DK 1.54\,m & 2019/05/23 & $I$ & $\# 1$ &-48.08 $\pm$ 40.88 &19.86 $\pm$ 26.27 & 9.79 $\pm$ 18.96 &0.46 $\pm$ 0.25 &4269 $\pm$ 620\\
DK 1.54\,m & 2019/06/07 & $I$ & $\# 1$ &24.89 $\pm$ 2.97 &11.90 $\pm$ 13.57 &5.50 $\pm$ 10.60 &0.51 $\pm$ 0.23 &4390 $\pm$ 546\\
DK 1.54\,m & 2021/05/23 & $I$ & $\# 1$ &0.70 $\pm$ 1.77 &24.52 $\pm$ 12.49 &16.91 $\pm$ 8.66 &0.77 $\pm$ 0.15 &4866 $\pm$ 336\\
DK 1.54\,m & 2022/05/13 & $I$ & $\# 1$ &2.38 $\pm$ 5.81 &15.34 $\pm$ 13.18 &9.68 $\pm$ 6.44 & 0.71 $\pm$ 0.22& 4762 $\pm$ 464\\  [2pt]
TESS S10 & $\#$ 2 & & $\# 1$ & 32.59 $\pm$ 43.65 & 20.28 $\pm$ 53.02& 18.35 $\pm$ 34.91& 0.60 $\pm$ 0.28 &  4575 $\pm$ 598\\
TESS S10 & $\#$ 4 & & $\# 1$ & 31.79 $\pm$ 52.82 &33.34 $\pm$ 54.25 &38.50 $\pm$ 48.00 &0.79 $\pm$ 0.28 & 4905 $\pm$ 539\\
TESS S10 & $\#$ 14~~ & & $\# 1$ & -5.91 $\pm$ 5.15 & 18.74 $\pm$ 13.47 & 22.66 $\pm$ 6.89 & 0.72 $\pm$ 0.10 & 4789 $\pm$ 250\\
TESS S10 & $\#$ 15~~ & & $\# 1$ & 16.88 $\pm$ 12.56 & 14.98 $\pm$ 27.90 & 7.15 $\pm$ 10.83 & 0.23 $\pm$ 0.27 & 3698 $\pm$ 920\\
TESS S36 & $\#$ 6 & & $\# 1$ &-39.04 $\pm$ 51.64 & 35.41 $\pm$ 53.42 & 37.95 $\pm$ 44.08  & 0.73 $\pm$ 0.29 & 4806 $\pm$ 571\\
TESS S36 & $\#$ 7 & & $\# 1$ & -14.43 $\pm$ 10.14 & 28.32 $\pm$ 28.73 & 16.76 $\pm$ 12.23 & 0.45 $\pm$ 0.24 & 4272 $\pm$ 589\\
TESS S36 & $\#$ 8 & & $\# 1$ & 8.94 $\pm$ 9.21 & 32.43 $\pm$ 33.01 & 23.49 $\pm$ 16.39 & 0.64 $\pm$ 0.22 & 4648 $\pm$ 474\\
TESS S36 & $\#$ 9 & 
&  
$\begin{tabular}{c}
$\# 1$ \\
$\# 2$ \\
\end{tabular}$
&  
$\begin{tabular}{c}
-15.82 $\pm$ 25.76 \\
-9.57 $\pm$ 48.03 \\
\end{tabular}$
&  
$\begin{tabular}{c}
24.39 $\pm$ 27.26 \\
72.92 $\pm$ 53.11 \\
\end{tabular}$
&  
$\begin{tabular}{c}
41.89 $\pm$ 25.78 \\
51.48 $\pm$ 33.69 \\
\end{tabular}$
&  
$\begin{tabular}{c}
0.76 $\pm$ 0.14 \\
0.76 $\pm$ 0.27 \\
\end{tabular}$
&  
$\begin{tabular}{c}
4856 $\pm$ 314 \\
4856 $\pm$ 529 \\
\end{tabular}$ \\
TESS S36 & $\#$ 10~~ & 
&  
$\begin{tabular}{c}
$\# 1$ \\
$\# 2$ \\
\end{tabular}$
&  
$\begin{tabular}{c}
9.47 $\pm$ 6.93 \\
51.09 $\pm$ 52.14 \\
\end{tabular}$
&  
$\begin{tabular}{c}
21.76 $\pm$ 15.68 \\
27.66 $\pm$ 52.45 \\
\end{tabular}$
&  
$\begin{tabular}{c}
13.18 $\pm$ 7.45 \\
30.40 $\pm$ 23.19 \\
\end{tabular}$
&  
$\begin{tabular}{c}
0.48$\pm$ 0.22 \\
0.51 $\pm$ 0.27 \\
\end{tabular}$
&  
$\begin{tabular}{c}
4337 $\pm$ 530 \\
4399 $\pm$ 620 \\
\end{tabular}$ \\
TESS S36 & $\#$ 11~~ & & $\# 1$ & 32.11 $\pm$ 42.00 & 20.37 $\pm$ 48.72 & 23.13 $\pm$ 31.48 & 0.77 $\pm$ 0.23 & 4872 $\pm$ 462\\
TESS S63 & $\#$ 3~~ & & $\# 1$ & 0.61 $\pm$ 44.00 & 16.92 $\pm$ 55.22 & 17.12 $\pm$ 46.76 & 0.70 $\pm$ 0.28 & 4755 $\pm$ 490\\
TESS S63 & $\#$ 9~~ & & $\# 1$ & -7.79 $\pm$ 32.00 & 17.57 $\pm$ 52.35 & 12.70 $\pm$ 38.57 & 0.43 $\pm$ 0.28 & 4229 $\pm$ 631\\
TESS S63 & $\#$ 11~~ & & $\# 1$ & 10.61 $\pm$ 50.86 & 24.02 $\pm$ 53.08 & 11.79 $\pm$ 41.32 & 0.62 $\pm$ 0.30 & 4613 $\pm$ 557\\
TESS S63 & $\#$ 13~~ & & $\# 1$ & -15.12 $\pm$ 40.22 & 18.75 $\pm$ 54.57 & 11.25 $\pm$ 39.66 & 0.56 $\pm$ 0.29 & 4499 $\pm$ 567\\
TESS S63 & $\#$ 14~~ & & $\# 1$ & 5.93 $\pm$ 54.84 & 22.41 $\pm$ 52.14 & 14.60 $\pm$ 41.39 & 0.60 $\pm$ 0.29 & 4575 $\pm$ 548\\
TESS S63 & $\#$ 15~~ & & $\# 1$ & 33.04 $\pm$ 49.80 & 16.93 $\pm$ 56.49 & 31.61 $\pm$ 46.23 & 0.76 $\pm$ 0.28 & 4857 $\pm$ 471\\
\hline
\multicolumn{9}{c}{\footnotesize  Notes: The values in the last column were obtained using Eq.~(\ref{eqn:contrast}).}
\end{tabular}
}
\end{table*}

%------------------------------
\section{Physical and starspot properties of the HATS-2 exoplanetary system}
\label{sec:prop}
%------------------------------
%------------------------------
\subsection{Stellar and planetary parameters}
%------------------------------
To estimate the main physical parameters of the HATS-2 system, we used the code \textsc{jktabsdim} \footnote{https://www.astro.keele.ac.uk/jkt/codes/jktabsdim.html} \citep{2009MNRAS.396.1023S}. The code makes use of the spectroscopic parameters, the stellar radial-velocity (RV) semi-amplitude ($K_{\star}=268.9 \pm 29.0$\,m\,s$^{-1}$; \citealt{2013A&A...558A..55M}) and a set of theoretical stellar models. In particular, \textsc{jktabsdim} uses tabulations from the Claret \citep{Claret}, Y$^2$ \citep{Demarque}, and Padova \citep{Girardi} models.  \textsc{jktabsdim} works in such a way that the velocity amplitude of the planet is iteratively modified to maximise the agreement between the best additional constraint derived from different theoretical models and the observed values of the stellar density from transit, effective temperature and metallicity \citep[$T_{\rm eff}=5227 \pm 95$ and ${\rm [Fe/H]}=0.15 \pm 0.05$;][]{2013A&A...558A..55M} . A wide range of possible ages for the parent star is also considered during this process. The code returns different estimates for each of the output parameters, one for each set of theoretical models. The unweighted means are considered the final values of the parameters. The systematic uncertainties, which were caused by the use of theoretical models, were estimated, while the statistical uncertainties were propagated from the uncertainties of the input parameters. It is worth to underline that \textsc{jktabsdim} does not use Gaia parallax, any apparent magnitudes, or interstellar extinction. As a consistency check, we calculated the distance to the system using our measurements of the stellar radius and effective temperature, apparent magnitudes in the optical ($BV$) and in the infrared (2MASS $JHK$), and the surface brightness relations from \cite{kervella}. An interstellar reddening of $E(B-V) = 0.08 \pm 0.03$\,mag was needed to align the distances in the optical and infrared passbands, and the distance found in the $K$-band is $332 \pm 9$\,pc. The Gaia DR3 distance of $337 \pm 2$\,pc is in good agreement with this value.

The final values of the parameters are listed in Table \ref{table:jktabsdim} and are in good agreement with those from \citet{2013A&A...558A..55M}. The only parameter that differs (although only at $\sim 1.5 \,\sigma$) is the planetary radius, which appears smaller in size. It is also possible to compare the final values with the ones obtained considering a weighted mean of the results of only the curves that do not present anomalies. The difference between these values can be considered as a first-order approximation of the systematic error introduced by the modelling of the starspots, which due to the low data quality, is within the 1-sigma uncertainties. In this way we derived that the systematic error is 0.85 $\%$ regarding the sum of the radii and 1.3 $\%$ regarding the ratio of the radii.

%\begin{table}
%\caption{Photometric properties of the HATS-2 system derived by fitting the transit light curves with the PRISM+GEMC and JKTEBOP codes. The values reported in the third column are the weighted means of the results for the individual data sets and are compared with those from the discovery paper.}
%\label{table:fit}
%\centering
%\resizebox{\hsize}{!}{
%\setlength{\tabcolsep}{12pt}
%\begin{tabular}{c c c c}
%\hline \hline 
%Parameter & Symbol & Weighted value & \citet{2013A&A...558A..55M}  \\ [1ex]
%\hline
%Sum of the radii   & $r_{\star} + r_{\rm p}$ & $0.2013 \pm 0.0038$  \\
%Ratio of the radii & $k$ & $0.1295 \pm 0.0013$ & $0.1335 \pm 0.0010$ \\
%Inclination        & $i\, (^{\circ})$ & $88.06 \pm 0.50$ & $87.2 \pm 0.7$\\
%Scaled stellar radius   & $r_{\star}$ & $0.1782 \pm 0.0035$ \\
%Scaled planetary radius & $r_{\rm p}$ & $0.02308 \pm 0.00051$ \\
%\hline
%\end{tabular}
%}
%\end{table}

\begin{table*}[h]
\caption{Median values and $68\%$ confidence intervals of the physical and orbital properties of the HATS-2 system, found using JKTABSDIM.}
\label{table:jktabsdim}
%\resizebox{\hsize}{!}{    
\centering      
\begin{tabular}{l c c c c}
\hline \hline\\[-6pt]%
Parameter & Symbol & Unit & This work & Mohler-Fischer et al. (2013)  \\ [1ex]
\hline\\[-6pt]%
Stellar mass & $M_{\star}$ &  $M_{\sun}$ & $0.904 \pm 0.051 \pm 0.010$ & $0.882 \pm 0.037$ \\ [1ex]
Stellar radius & $R_{\star}$ & $R_{\sun}$ & 0.876 $\pm$ 0.020 $\pm$ 0.003 & 0.898 $\pm$ 0.019 \\[1ex]
Stellar logarithmic surface gravity & $\log{g_{\star}}$ & cgs & 4.510 $\pm$ 0.012 $\pm$ 0.002 & $4.48 \pm 0.02$ \\[1ex]
Stellar density & $\rho_{\star}$ & $\rho_{\sun}$ & $1.346 \pm 0.028$ & - \\[1ex]
Stellar Age & $\tau_{\star}$ & Gyr & ${7.2}^{+3.7\,+2.9}_{-6.3\,-5.1}$  & $9.7 \pm 2.9$ \\[1ex]
Stellar Rotational Period & $P_{\rm rot}$ & day & $22.46 \pm 5.20$ & $30.32 \pm 10.13$ \\[1ex]
Planetary mass & $M_{\rm p}$ & $M_{\rm Jup}$ & 1.37 $\pm$ 0.16 $\pm$ 0.01 & 1.345 $\pm$ 0.150 \\[1ex]
Planetary radius & $R_{\rm p}$ & $R_{ \rm Jup}$ & 1.122 $\pm$ 0.024 $\pm$ 0.004 & 1.168 $\pm$ 0.030 \\[1ex]
Planetary surface gravity & $g_{\rm p}$ & m\,s$^{-2}$ & 27.0 $\pm$ 2.9 & 24.6 $\pm$ 2.8  \\[1ex]
Planetary density & $\rho_{\rm p}$ & $\rho_{\rm Jup}$ & 0.91 $\pm$ 0.10 & 0.79 $\pm$ 0.11\\[1ex]
Planetary equilibrium temperature & $T_{\rm eq}$ & K& 1550 $\pm$ 29 & 1577 $\pm$ 31\\  [1ex]
Orbital inclination &$i$ & $^{\circ}$ & 87.83 $\pm$ 0.18 & 87.2 $\pm$ 0.7\\ [1ex]
Semi-major axis & $a$ & au & 0.02317 $\pm$ 0.00043 $\pm$ 0.00009 & 0.0230 $\pm$ 0.0003\\[1ex]	
Inclination angle of the stellar spin  & $\sin i_{\star}$ &   & 0.76 $\pm$ 0.31 & - \\[1ex]
Sky-projected spin-orbit angle & $\left| \lambda \right|$ & $^{\circ}$ & 2.72 $\pm$ 17.84 & 8 $\pm$ 8 \\[1ex]
True spin-orbit angle & $\psi$ &  $^{\circ}$ & $38.49 \pm 27.13$ & - \\[1ex]
\hline
\multicolumn{5}{c}{\footnotesize Notes: Results from \citet{2013A&A...558A..55M} are also reported for comparison. Where two error bars are }\\
\multicolumn{5}{c}{\footnotesize given, the first refers to the statistical uncertainties, while the second to the systematic errors. Other derived }\\
\multicolumn{5}{c}{\footnotesize parameters, i.e. the stellar rotational period, the sky-projected obliquity, the inclination angle of the stellar spin,}\\
\multicolumn{5}{c}{\footnotesize  and the true obliquity are also listed.}\\
\end{tabular}
%}
\end{table*} 

%------------------------------
\subsection{Starspot analysis}
%------------------------------
\begin{figure*}[h]
%	\centering
%	\includegraphics[width=0.8\linewidth]{contrast_sub}
\resizebox{\hsize}{!}{\includegraphics{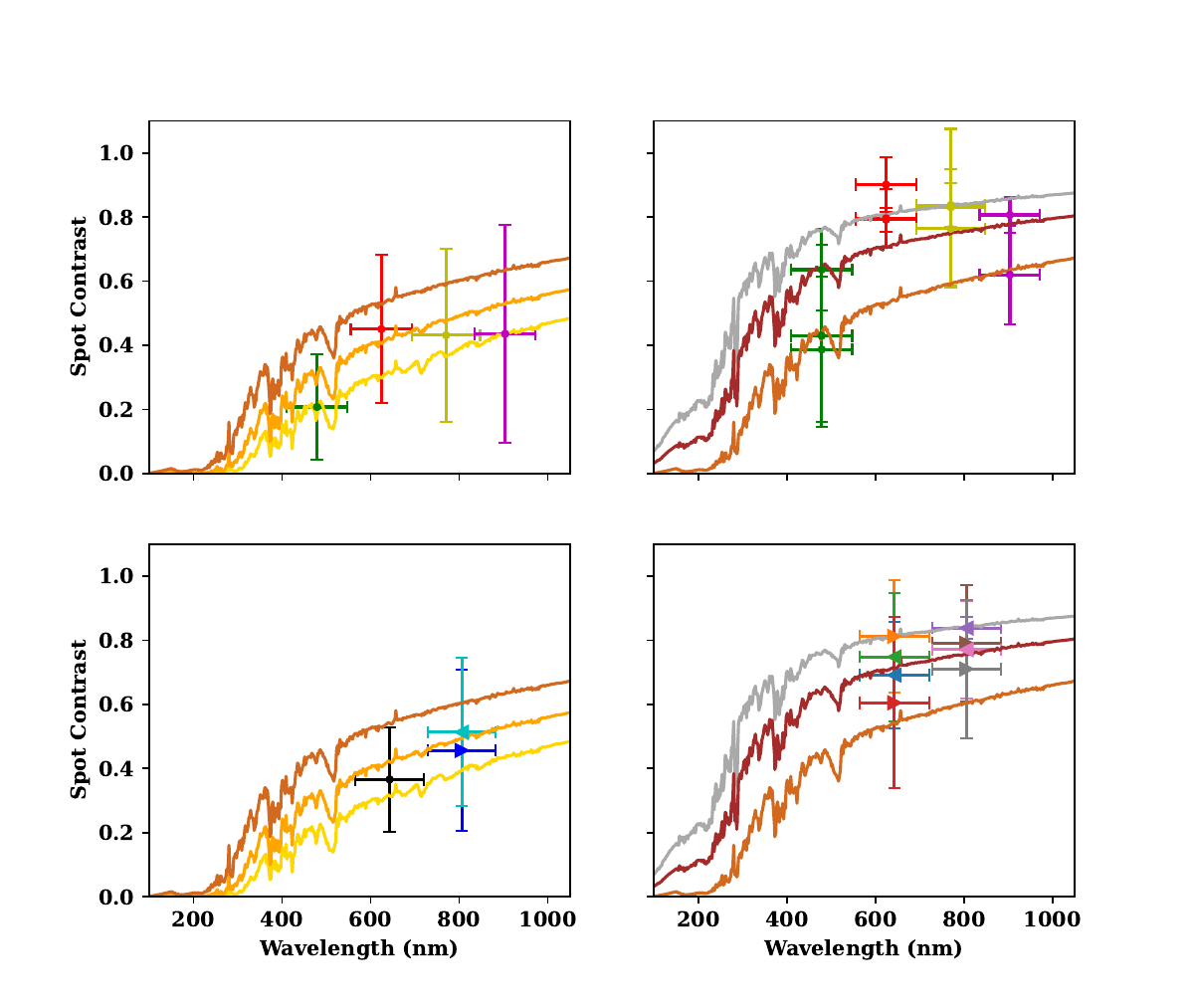}}
\caption{Variation in the starspot contrast with wavelength. The vertical bars represent the errors in the measurements and the horizontal bars show the FWHM transmission of the passbands used. In the upper panels, the colour of the data points is in agreement with Fig.~\ref{fig:grond}. In the top left panel, the data taken on 2012 February 28 with the GROND multi-band camera are represented together with the ATLAS9 \citep{1979ApJS...40....1K} stellar atmospherical models computed considering a stellar surface at 5227\,K and a starspot at 4200 and 4400\,K. In the top right panel, the data taken on 2012 June 1st with the GROND multi-band camera are represented together with the ATLAS9 stellar atmospherical models computed considering a stellar surface at 5227\,K and a starspot at 4600, 4850, and 4980\,K. In the bottom panels, the points represent the contrast values obtained using the Danish 1.54\,m transits (as explained in the plot legend), while the curves are the same ATLAS9 models used in the upper panels.}
\label{fig:contrastsub}
\end{figure*}

Having modelled the HATS-2 transit light curves containing starspot anomalies with the PRISM-GEMC codes, we were able to derive the best-fitting values for the position of the starspots on the stellar disc, beyond their angular size and contrast. These estimations are reported in Table~\ref{tab:fitsp}, in whose last column the reader can also find the starspot-temperature values, which were derived considering the stellar photosphere and starspots as black bodies and using equation 1 from \citet{2003ApJ...585L.147S}, that is
\begin{equation}
\rho= \frac{B_{\nu} (T_{\rm spot}) }{B_{\nu} (T_{\rm eff})}= 
\frac
{{\rm exp}\left(h\nu/k_{\rm B} T_{\rm eff}-1\right)}
{{\rm exp}\left(h\nu/k_{\rm B} T_{\rm spot}-1\right)}
%\frac{e^{\frac{h\nu}{k_BT_{\star} }}-1}{e^{\frac{h\nu}{k_BT_{\rm spot} }}-1}
\, ,
\label{eqn:contrast}
\end{equation}
where $\nu$ is the central frequency of each filter band\footnote{http://ulisse.pd.astro.it/Astro/ADPS/}, $h$ is the Planck constant, while $k_{\rm B}$ is the Boltzmann constant. 
All the values are consistent with starspots cooler than the stellar photosphere and lie between 4000 and 5000\,K, within the uncertainties. 

The only exception is related to the first anomaly on the $g^{\prime}$ light curve of the transit observed on 2012 June 1st, which was identified by \citet{2013A&A...558A..55M} as a plage (see the top right-hand panel in Fig.~\ref{fig:grond}). In the case of the Sun, plages are usually visible through H-$\alpha$  
or calcium (Ca) K line wavelengths by using appropriate filters. 
In K-type stars, such as HATS-2, the Ca\,II lines are much stronger than the H-$\alpha$ ones and fall in the wavelength transmission range defined by the $g^{\prime}$ filter, $\lambda=478.8 \pm 137.9$\,nm. We, therefore, agree with the interpretation of \citet{2013A&A...558A..55M} that the anomaly could be caused by a plage connected with the following starspot.

In general, we noted that the angular sizes of the starspots vary from a minimum of $2.9^{\circ}$ to a maximum of $40^{\circ}-50^{\circ}$. Considering that an angular radius of $90^{\circ}$ covers half of the stellar hemisphere, the starspots we have found cover from a minimum of $1.6\%$ to a maximum of $20\%-30\%$ of the stellar disc.
%, and so from $0.8\%$ to $11\%-14\%$ of the total stellar %surface. 
As in the case of, e.g. TrEs-1 and WASP-19, our measurements are similar to those found for other G-type and K-type stars \citep{2009A&A...494..391R,2013MNRAS.436....2M} and are in good agreement with the sizes of very large sunspots  \citep{sun}. However, we stress that some of the starspots, which we have detected and modelled with circular shapes, might actually be a group of starspots, according to the Zurich and McIntosh classifications \citep{mcintosh}. 

Since starspots are darker in the ultraviolet than in the infrared and we observed transits through different filters, we can check if the starspot contrast varies as expected as a function of the wavelength. In Fig.~\ref{fig:contrastsub} we compare the starspot contrasts estimated by PRISM+GEMC and reported in Table~\ref{tab:fitsp} with theoretical expectations.
Using ATLAS9 atmospheric models \citep{1979ApJS...40....1K}, we modelled a stellar photosphere of $5227$\,K and starspots with five different temperature $4200$\,K, $4400$\,K, $4600$\,K, $4850$\,K and $4980$\,K. Although the error bars of the starspot contrasts are quite large, the trend in each panel of Fig.~\ref{fig:contrastsub} is that for which the starspots are brighter in the redder passbands than in the bluer one, for both all three complete simultaneous GROND multi-band observations and the two-colours observations with the Danish telescope.

%------------------------------
\subsection{Spin-orbit alignment}
%------------------------------
We investigated the possibility that some of the observed starspot-crossing events have been caused by the same starspot because, then, as explained in Sect.~\ref{Sec:introduction}, the spin-orbit alignment of the planetary system can be derived. 

\citet{2013A&A...558A..55M} already argued that the anomalies observed during the 2012 February 28 and 2012 June 1st transits could have been caused by the same starspot. Studying the values reported in Table~\ref{tab:fitsp}, we note that, for these two transits, the differences in the starspot latitudes in all the GROND bands fall within $1\sigma$ but their angular sizes are not similar. Hence, it is possible that we are dealing with a starspot complex that evolved following the Zurich classification. However, the two transits are separated by 94 days and, using empirical laws, \citet{2013A&A...558A..55M} estimated that a group of starspots with this size could have a lifetime of $\simeq 130$\,days. With these caveats, the authors estimated a sky-projected obliquity of $8^{\circ} \pm 8^{\circ}$ and a rotational period of $31\pm 10$\,days. 

\begin{figure}%
\centering
{{\includegraphics[width=4.1cm]{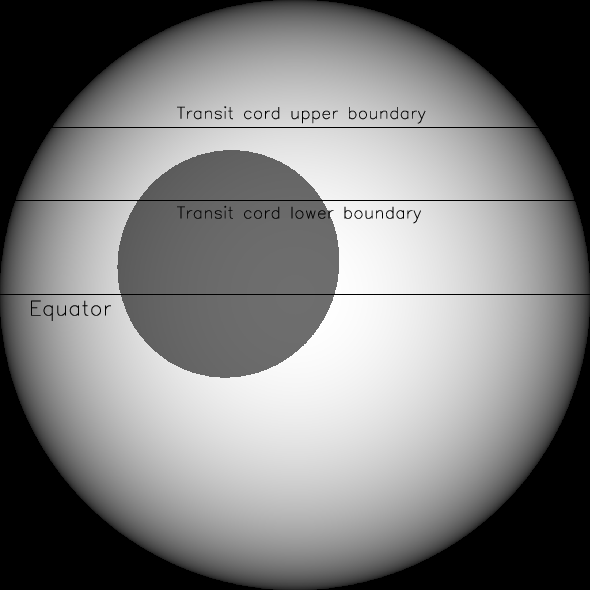}}}%
\qquad
{{\includegraphics[width=4.1cm]{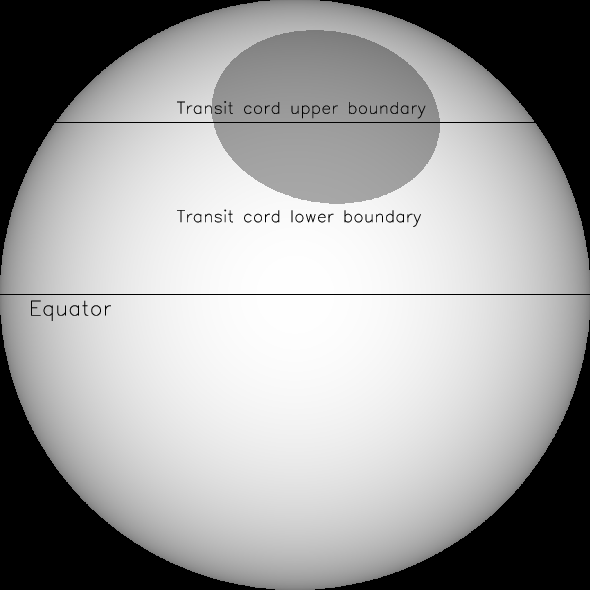}}}%
\caption{Representation of the stellar disc, starspot position, and transit chord for the two consecutive transit events, $\#7$ (left panel) and $\#8$ (right panel) with starspot crossings, obtained using PRISM/GEMC. The grey scale of each starspot is related to its contrast. The starspot properties refer to the best-fitting models in Fig. \ref{fig:tess3} (see Appendix).}%
\label{fig:sp}%
\end{figure}

Using the new data presented in this work, we identified eleven consecutive starspot-crossing events, separated by less than 5 days, and for which the relative latitudes and angular sizes differences fall within $1\sigma$, referring to the values listed in Table~\ref{tab:fitsp}. 10 out of 11 events have been identified in the TESS photometric data. Of particular interest are the consecutive transits that we labelled with $\#7$ and $\#8$ from TESS Sector 36 (see Fig.~\ref{fig:sp}). Investigating the light curves (see Fig.~\ref{fig:tess3}), we note that the anomalies that appear in the transits $\#7$ and $\#8$ have similar amplitudes and duration, whereas the time of the maximum of the anomalies changes with the increase of the orbital phase. These facts suggest that the anomalies are due to the same starspot rotating around the surface of the star.
%
%\begin{figure}[h]
%\centering
%\includegraphics[width=0.9\linewidth]%{starspot.pdf}
%\caption{Representation of the stellar disc, %star-spot position, and transit chord for the %transit events with star-spot crossings, %obtained using PRISM/GEMC. The grey scale of %each starspot is related to its contrast. The %starspot properties refer to the best-fitting %models in Fig. \ref{fig:tess3} (see %Appendix). Labels indicate the observational %ID.}
%\label{fig:sp}
%\end{figure}
%
This hypothesis is also supported by the modelling results from PRISM/GEMC. Indeed, the best-fitting circular starspots for the transit events $\#7$ and $\#8$ have similar sizes and contrasts (see  Table~\ref{tab:fitsp}). The transit designed as number $\#6$ also supports this hypothesis, although its small size does not allow us to constrain the size of the starspot with the required accuracy.

%Moreover, Fig.~\ref{fig:678sp} shows very clearly that %the HATS-2 system may be characterized by a low but non-%zero angle between the direction of the stellar %rotation and the planet's orbital plane. 
The other TESS consecutive transit events were identified as $\#$14 and $\#$15 from Sector 10 (Fig.~\ref{fig:tess1}), as $\#$9, $\#$10, and $\#$11 from Sector 36 (Fig.~\ref{fig:tess4}), and as $\#$13, $\#$14, and $\#$15 from Sector 63 (Fig.~\ref{fig:tess6}). 

The two consecutive starspot-crossing events, which were identified using our data from ground-based telescopes, are those from the Danish 1.54\,m telescope observed on 2019 May 19 and 2019 May 23 (see Fig.~\ref{fig:dan}). 

\begin{figure}[h]
\centering
\includegraphics[width=\hsize]{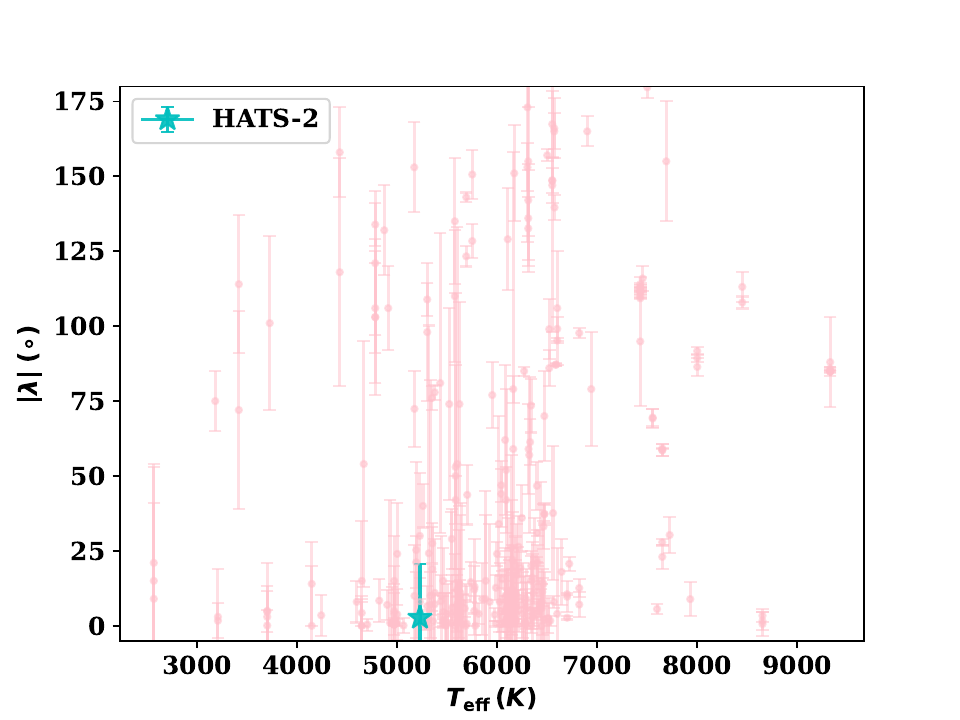}
\caption{Sky-projected obliquity as a function of the host stars' effective temperature for an ensemble of exoplanetary systems. The star is the value of the sky-projected obliquity of the HATS-2 system found in this work, while the dots are the values from the TEPCAT catalogue. }
\label{fig:hats2_obl}
\end{figure}

Using the values listed in Table \ref{tab:fitsp}, it is possible to calculate the values of the sky-projected obliquity, $\lambda$, and the stellar rotational period $P_{\rm rot}$ for each of the identified events by using the following equations:
\begin{equation}
   \left| \lambda \right| = \arctan{\left( \left| \frac{\sin \theta_2 - \sin \theta_1}{\cos \theta_2 \sin \phi_2 - \cos \theta_1 \sin \phi_1} \right| \right)}\, ,
    \label{eqn:lambda}
\end{equation}
\begin{equation}
    P_{\rm rot} = \frac{2\pi \Delta t}{\sqrt{D}}\, ,
    \label{eqn:prot}
\end{equation}
\begin{multline}
\label{eqn:d}
D=  2\left( 1 - \sin \theta_1 \sin \theta_2 - \cos \theta_1 \cos \theta_2 \sin \phi_1 \sin \phi_2 \right) \\ + \cos^2 \theta_2 \left[ \sin^2 \phi_2 -1 \right] + \cos^2 \theta_1 \left[ \sin^2 \phi_1 -1 \right]\,,
\end{multline}
where $\theta_{1,2}$ and $\phi_{1,2}$ are the latitudes and longitudes of the two consecutive starspots and $\Delta t$ is the time interval between the two starspot-crossing events. The quantity $\sqrt{D}$ represents the arc length on the stellar photosphere between the two positions of the starspot. The mathematical derivation of these equations is given in the Appendix. The values of $\lambda$ and $P_{\rm rot}$ are reported in Table~{\ref{tab:lambda}} for each event. The large uncertainties associated with the $\left| \lambda \right|$ and $P_{\rm rot}$ values derived from some of the TESS light curves must be once again intended, as explained in Section \ref{sec:light_curve_analysis}, as a consequence of the poor quality (high scatter) of the data. Our final estimations were computed by performing a weighted mean, obtaining {$\left| \lambda \right|=2^{\circ}.72 \pm 17^{\circ}.84$ and $P_{\rm rot}=22.46 \pm 5.20$\,days}. This value is consistent with the value of $P_{\rm rot, \, TESS}=21.44$\,days, obtained with a FAP$<0.1\,\%$ (False Alarm Probability) performing a GLS (Generalised Lomb-Scargle Periodgram, \citealt{lsp}) analysis of the SAP flux from all the three sectors of the TESS Mission, see Fig. \ref{fig:gls}. The estimation of the sky-projected obliquity resulted consistent with zero degrees and, therefore, with a well-aligned system. It is represented in Fig.~\ref{fig:hats2_obl} as a function of the parent-star effective temperature, together with a sample of values taken from the TEPCat catalogue\footnote{The TEPCat catalogue is available at \url{http://www.astro.keele.ac.uk/jkt/tepcat} \citep{2011MNRAS.417.2166S}.}.
\begin{table} 
\caption{All the spot-crossing events identified over two transits in this work.}
\label{tab:lambda}
\resizebox{\hsize}{!}{  
\centering
\begin{tabular}{l c c c}
\hline \hline    
Telescope & Consecutive Events & $\left| \lambda \right|$ ($^{\circ}$) & $P_{\rm rot}$ (days)\\
\hline
\hline
\\ [-8pt]
Danish& May 19-23, 2019       & $0.70  \pm~~ 22.0   $& $22.01 \pm~~ 7.06~$\\
TESS s10 & $\#$14 and $\#$15  & $9.40  \pm  76.20   $& $22.21 \pm  14.32 $\\
TESS s36 & $\#$6 and $\#$7    & $20 \pm  176~~ $& $27.3 \pm  49.5 $\\
TESS s36 & $\#$6 and $\#$8    & $3.83  \pm  81.47   $& $26.34 \pm  26.8 $\\
TESS s36 & $\#$7 and $\#$8    & $10.01 \pm  104.00~~ $& $23.90 \pm  16.33 $\\
TESS s36 & ~~$\#$9 and $\#$10 & $6.01  \pm  71.80   $& $21.10 \pm  21.41 $\\
TESS s36 & ~~$\#$9 and $\#$11 & $4.97  \pm  68.53   $& $22.71 \pm  22.30 $\\
TESS s36 & $\#$10 and $\#$11  & $3.75  \pm  136.63   $& $24.57 \pm  44.57 $\\ 
TESS s63 & $\#$13 and $\#$14 & 9.90 $\pm$ 201.64 & 24.47 $\pm$ 77.67 \\
TESS s63 & $\#$13 and $\#$15 & 2.25 $\pm$ 97.00 & 22.12 $\pm$ 27.82 \\
TESS s63 & $\#$14 and $\#$15 & 11.93 $\pm$ 161.79 & 19.53 $\pm$ 52.54 \\[2pt]
& & \textbf{2.72 $\pm$ 17.84}~~ & \textbf{22.46 $\pm$ 5.20~~} \\
\hline
\multicolumn{4}{c}{\footnotesize  Notes: In the table are listed the sky-projected and stellar rotational period }\\
\multicolumn{4}{c}{\footnotesize  values obtained using Eq. (\ref{eqn:lambda}) and (\ref{eqn:prot}). The final values were }\\
\multicolumn{4}{c}{\footnotesize  calculated by a weighted mean.}\\
\end{tabular}
}
\end{table}	

Considering that HATS-2 has $T_{\rm eff}=5227\pm 95$ K, our result is in agreement with the trend for which the orbits of the hot Jupiters hosted around cool dwarf stars tend to present a low obliquity \citep{2010ApJ...718L.145W}. Moreover, this can be an indication that HATS-2\,b has reached its actual position through a Type II migration, or that tidal interactions with its parent star damped a non-zero obliquity caused by Lidov-Kozai oscillations induced by a distant companion \citep{2022ASSL..466....3R}. The second scenario is preferable as HATS-2 is a quite old ($\sim 7$\,Gyr) K-type star, so both the age and the convective envelope enhance the tidal dissipation efficiency. The analysis of the TTVs, which is discussed in the next section, will provide other insights to discriminate between the two formation scenarios.

Knowing the rotational period of the parent star and using $(v \sin{i_{\star}})$ from \cite{2013A&A...558A..55M}, we determined the sine of inclination of the stellar-rotation axis by means of
\begin{equation}
	\sin{i_{\star}} = P_{\rm rot} \frac{(v \sin{i_{\star}})}{2\pi R_{\star}}=0.76 \pm 0.31\,.
	\label{eqn:sini}
\end{equation}
Then, using the following formula \citep{2007AJ....133.1828W}
\begin{equation}
	\cos{\psi} = \cos{i} \cos{{i_{\star}}} + \sin{i} \sin{{i_{\star}}} \cos{\lambda},
	\label{eqn:psi}
\end{equation}
we derived the true orbital obliquity of HATS-2\,b, {which resulted to be $\psi=38^{\circ}.49 \pm 27^{\circ}.13$}. This measurement excludes that HATS-2 is in a pole-on configuration.

\section{Transit Timing Variations}
\label{sec:ttv}
%----------------------------------
\begin{table}
\caption{ Times of mid-transit for HATS-2\,b and relative O-C residuals.}
\label{table:eff}
\resizebox{\hsize}{!}{  
\centering
\begin{tabular}{c r r r}
\hline \hline \\[-6pt]
Telescope & Epoch & O (BJD$_{\rm TDB}$)~~~~~~~~~~~~~ & O-C (min)~~   \\ [1ex]
\hline\\[-6pt]
MPG & -1924.0   & 55985.731172 $\pm$ 0.000090   & -0.501 $\pm$ 0.130 \\
MPG & -1854.0   & 56080.520649 $\pm$ 0.000109   & -0.340 $\pm$ 0.157 \\
MPG & -1139.0   & 57048.727133 $\pm$ 0.000117   & 0.847 $\pm$ 0.168\\
DK & -809.0   & 57495.591314 $\pm$ 0.000161   & 0.891 $\pm$ 0.232\\
DK & -507.0   & 57904.539218 $\pm$ 0.000168   & 0.170 $\pm$ 0.241\\
DK & -476.0   & 57946.517242 $\pm$ 0.000175   & -0.007 $\pm$ 0.251\\
DK & -247.0   & 58256.613740 $\pm$ 0.000180   & -0.209 $\pm$ 0.259\\
DK & -233.0   & 58275.571993 $\pm$ 0.000189   & 0.338 $\pm$ 0.273\\
DK & -219.0   & 58294.529570 $\pm$ 0.000198   & -0.088 $\pm$ 0.285\\
TESS & -10.0   & 58577.543615 $\pm$ 0.000203   & 0.031 $\pm$ 0.292\\
TESS & 0.0   & 58591.084807 $\pm$ 0.000227   & -0.178 $\pm$ 0.327\\
DK & 24.0   & 58623.584665 $\pm$ 0.000282   & 0.752 $\pm$ 0.406\\
DK & 27.0   & 58627.646453 $\pm$ 0.000282   & -0.130 $\pm$ 0.406\\
DK & 38.0   & 58642.542160 $\pm$ 0.000301   & 0.204 $\pm$ 0.434\\
DSC/ExoClock & 468.0   & 59224.819071 $\pm$ 0.000317   & -0.682 $\pm$ 0.457\\
SSO/ExoClock &  500.0   & 59268.151393 $\pm$ 0.000332   & -0.623 $\pm$ 0.478\\
TESS &514.0   & 59287.109786 $\pm$ 0.000379   & 0.124 $\pm$ 0.546\\
TESS &526.0   & 59303.359703 $\pm$ 0.000383   & 0.574 $\pm$ 0.551\\
DSC/ExoClock & 535.0   & 59315.546858 $\pm$ 0.000383   & 0.503 $\pm$ 0.552\\
DSC/ExoClock & 538.0   & 59319.608897 $\pm$ 0.000403   & -0.018 $\pm$ 0.581\\
DK & 569.0   & 59361.587384 $\pm$ 0.000424   & 0.469 $\pm$ 0.610\\
DSC/ExoClock & 736.0   & 59587.726556 $\pm$ 0.000451   & -1.217 $\pm$ 0.650\\
DSC/ExoClock & 1010.0   & 59958.758793 $\pm$ 0.000461   & -1.823 $\pm$ 0.664\\
TESS &1054.0   & 60018.341413 $\pm$ 0.000484   & -0.767 $\pm$ 0.696\\
TESS &1058.0   & 60023.758037 $\pm$ 0.000507   & -0.639 $\pm$ 0.730\\
TESS &1063.0   & 60030.529038 $\pm$ 0.000584   & -0.162 $\pm$ 0.841\\
TESS &1068.0   & 60037.298643 $\pm$ 0.001085   & -1.693 $\pm$ 1.562\\
\hline
\multicolumn{4}{c}{\footnotesize  Notes: The values refer to Case ($b$), and are calculated using equation (\ref{eqn:eff2}). }\\
\end{tabular}
}
\end{table}

Thanks to the analysis of the light curves presented in the previous sections, we have a set of mid-transit times, which are listed in Table~\ref{table:eff}. The values derived from the GROND light curves refer to a weighted mean of the $T_0$ values obtained in each of the four optical filters, while the values derived from the TESS light curves refer to the result of the eight simultaneous fits with JKTEBOP. 

We also used a series of light curves from the ExoClock database \citep{kokori1} \footnote{\url{https://www.exoclock.space/database/planets/HATS-2b/}.}. In particular, we selected the most well-sampled (i.e. with well-sampled ingress, plateau and egress) and high-quality (i.e. low scatter; see the last column of Table \ref{tab:obs}) light curves from those presented by \citet{ed} and \citet{kokori2}. In detail, we selected three light curves collected at the Deep Sky Chile Observatory on 2021 January 10, 2021 April 11, and 2021 April 15 \citep{kokori2}, and one at the Sliding Spring Observatory on 2021 February 22 \citep{ed}. 

We also considered two unpublished light curves once more collected by amateur astronomers within the ExoClock collaboration; one was again observed with the Deep Sky Chile 40\,cm telescope, while the other with the Hakos Astro Farm 50\,cm telescope\footnote{\url{https://www.exoclock.space/database/observations_by_observer}.}.

%We also obtained two unpublished light curves via private communication with astronomers involved in the %ExoClock collaboration\footnote{F.B. and L.M. thank Yves Jorgen and Jean-Pascal Vignes, also co-authors of %this work}. 

All these extra light curves were fitted with the same procedures presented in the previous sections thanks to the JKTEBOP and PRISM+GEMC codes. The best-fitting models are presented in Fig.~\ref{fig:exoclock} together with the detrended light curves. The results are listed in Table \ref{tab:exoclockfit}. In the end, our final ensemble of mid-transit times covers a time baseline extending over more than 10 years. 
It is, therefore, possible to refine the ephemerides by considering various models for fitting the timings. 

We set the middle transit of the second half of the Sector 10 TESS data to be the zero$^{\rm th}$ epoch because most of its light curve does not show the presence of anomalies connected to starspots within the experimental uncertainties and has been estimated by a simultaneously fit with the `neighbour' transits. 

\begin{figure}[h]
\centering
\hspace*{-0.4cm}
\includegraphics[width=1.1\linewidth]{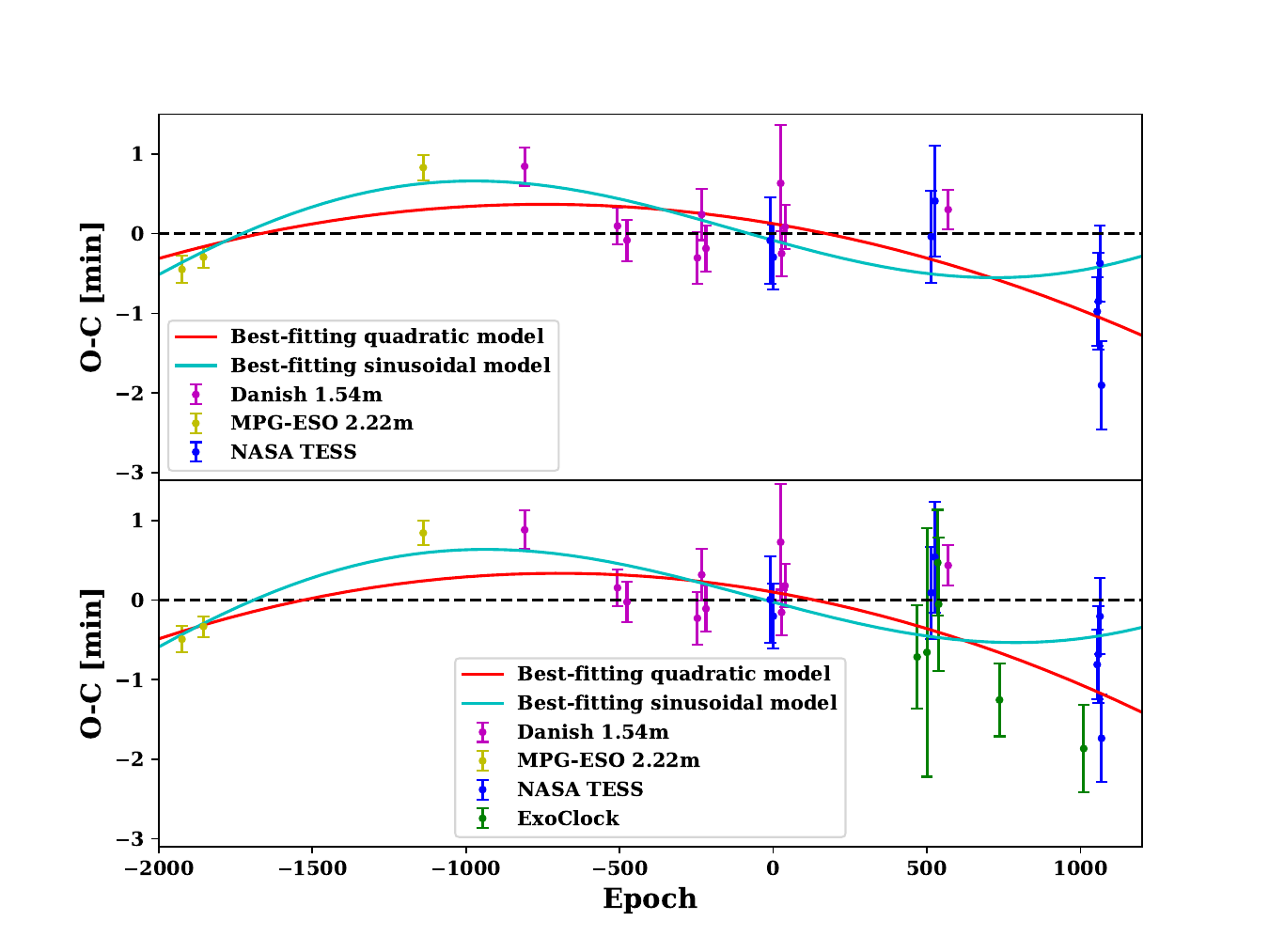}
\caption{{\it Top panel:} O-C residuals for HATS-2\,b in the case ($a$), superimposed with the best-fitting linear ephemeris model (black line), the best-fitting quadratic ephemeris model (red line) and the best-fitting cosinusoidal ephemeris model (cyan line). The purple points represent the times derived from the Danish 1.54\,m telescope observations, the yellow points the times derived from the MPG/ESO 2.2m telescope data, and the blue points the times derived from TESS data. {\it Bottom panel}: Same as the top panel but for case ($b$), i.e. also considering the data from the ExoClock database, which are represented as green points. }
\label{fig:hats2eff}
\end{figure}

We considered two different data sets: ($a$) fitting only GROND, DFOSC and TESS data without ExoClock times, and ($b$) fitting all the data listed in Table \ref{table:eff}. The first model that we tried is a simple straight line, $T_0^{\rm lin}=\kappa_0 +\kappa_1\times E$, which represents a linear ephemeris. In case ($a$), we obtained
\begin{equation}
	T_0^{\rm lin}= \, {\rm BJD_{TDB}}\, 58591.08501(11) + 1.35413385(9) \times E \,,
	\label{eqn:eff1}
\end{equation}    
while in case ($b$)
\begin{equation}
	T_0^{\rm lin}= \, {\rm BJD_{TDB}}\, 58591.084931(96) + 1.35413379(8) \times E \,.
	\label{eqn:eff2}
\end{equation}    
Here, $E$ represents the epoch of the transit and the bracketed quantities indicate the uncertainties in the preceding digits. Variations of the order of minutes are evident from the O-C residuals of both of the two fits shown in Fig.~\ref{fig:hats2eff}. This means that the linear ephemeris is a good approximation but seems to not exactly reproduce the dynamic state of the system. 

Since physical phenomena may hide in these variations, we also tried to fit the data with 
\begin{itemize}
\item[$\bullet$] a quadratic polynomial that represents a quadratic ephemeris and could be connected to a tidal-decay scenario
\begin{equation}
T_0^{\rm quad}= \kappa_0 + \kappa_1 \times E + \kappa_2\times E^2 \,;
\label{eqn:quadeff}
\end{equation}
\item[$\bullet$] a sinusoidal function that could be seen as a linear ephemeris periodically perturbed with an amplitude $A_{\rm TTV}$, a period $P_{\rm TTV}={2\pi}/{\Omega}$, and a phase $\varphi$. This ephemeris could be connected to an apsidal precession scenario or the presence of a planetary or sub-stellar companion (e.g. a brown dwarf)
\begin{equation}
T_0^{\rm sin}= \kappa_0 + \kappa_1\times E + A \sin \left(  \Omega \,E + \varphi  \right).
\label{eqn:sineff}	
\end{equation} 
\end{itemize}  
All the fits were performed using the {\texttt{lmfit}} \citep{lmfit} Python package. The results are listed in Table \ref{table:coeff} and their O-C residuals versus the best-fitting linear model are represented in Fig.~\ref{fig:hats2eff}, for both scenarios. The sinusoidal and quadratic ephemeris seems to fit better the data with respect to the linear one as confirmed by statistical indicators, such as the Reduced Chi-Squared $\chi_{\rm red}$, the Bayesian information criterion (BIC) and the Akaike information criterion (AIC), which are, respectively, defined as
\begin{equation}
	\chi_{\rm red}^2=\frac{\chi^2}{N_{\rm dof}}=\frac{\chi^2}{N-j } \; ,
	\label{eqn:chired}
\end{equation} 
\begin{equation}
	{\rm BIC} = N\ln \left( \frac{\chi^2}{N} \right) + N_{\Theta}\ln N\; ,
	\label{eqn:bic}
\end{equation} 
\begin{equation}
	{\rm AIC} = N\ln \left( \frac{\chi^2}{N} \right) + 2N_{\Theta}\; ,
	\label{eqn:aic}
\end{equation}
where $N$ is the number of data points and $N_{\Theta}$ is the number of parameters of the best-fitting model. $N_{\rm dof}$ is the number of degrees of freedom and usually is set as the number of data points $N$ minus the number of free parameters $N_{\Theta}$. The $\chi_{\rm red}^2$, BIC and AIC values for each of the considered ephemeris models and groups of data are listed in Table \ref{table:chired}. 

\begin{table*}
\caption{Best-fit parameters of the ephemeris models for the HATS-2\,b mid-transit time residuals. }
\label{table:coeff}
\resizebox{\hsize}{!}{  
\centering
\begin{tabular}{l c c c c c c}
\hline \hline \\ [-6pt]%
~~~Model & $\kappa_0$ & $\kappa_1$ & $\kappa_2$  & $A$ & $\Omega$ & $\varphi$ \\
& [days] & [days/epoch] & [days/epoch$^2$] & [days] & [rad/epoch] &  [rad]\\
\hline \\ [-6pt]%
\hline \\ [-6pt]%
& & & Data from professional telescopes -- Case ($a$) & & & \\
\hline \\ [-6pt]%
\hline \\ [-6pt]%
Linear     & $58591.08501 \pm 0.00011$ & $1.35413385 \pm 0.00000009$ & & & & \\
Quadratic  & $58591.08510 \pm 0.00009$ & $1.35413340 \pm 0.00000010$ & $(-3.01 \pm 0.85) \times 10^{-10}$  & & & \\ 
Sinusoidal & $58591.08510 \pm 0.00010$ & $1.35413424 \pm 0.00000016$ & & $(7.98 \pm 1.04) \times 10^{-4}$ & $0.00144 \pm 0.00018$ & $3.33 \pm 0.19$ \\
\hline \\ [-6pt]%
\hline \\ [-6pt]%
& & & Data from professional telescopes + ExoClock -- Case ($b$) & & & \\
\hline \\ [-6pt]%
\hline \\
Linear     & $58591.084931 \pm 0.000096$ & $1.35413379 \pm 0.00000008$ & & & & \\
Quadratic  & $58591.085000 \pm 0.000070$ & $1.35413330 \pm 0.00000010$ & $(-3.37 \pm 0.75) \times 10^{-10}$  & & & \\ 
Sinusoidal & $58591.085000 \pm 0.000200$ & $1.35413420 \pm 0.00000096$ & & $(8.6 \pm 7.0) \times 10^{-4}$ & $0.00137 \pm 0.00066$  & $3.25 \pm 0.23$ \\
\hline 
\end{tabular}
}
\end{table*}
\begin{table}
\caption{Goodness of fit of the three ephemeris models.}
\label{table:chired}
\resizebox{\hsize}{!}{  
\centering
\begin{tabular}{c c c c c c c}
\hline \hline \\ [-6pt]%
Data set& Model & $N_{\Theta}$ & $\chi^2$ & $\chi_{\rm red}^2$ & BIC& AIC\\ [1ex]
\hline \\ [-6pt]%
Case ($a$) & Linear Model & 2 & 72.54 & 3.82 & 32.12 & 30.03 \\ [1ex]
&Quadratic Model & 3 & 39.66 & 2.20 & 22.49 & 19.35\\ [1ex]
&Sinusoidal Model & 5 & 30.91 & 1.93 & 23.34 & 18.12\\ [1ex]
\hline \\ [-6pt]%
Case ($b$) & Linear Model & 2 & 85.83 & 3.30 & 38.03 & 35.36 \\ [1ex]
&Quadratic Model & 3 & 44.48 & 1.78 & 22.95 & 18.96\\ [1ex]
&Sinusoidal Model & 5 & 34.96 & 1.52 & 22.87 & 16.22\\ [1ex]
\hline
\end{tabular}
}
\end{table}

All the $\chi_{\rm red}^2$ values listed in Table \ref{table:chired} are larger than unity. This is not a sign of a bad fit but is in line with our previous TTV analysis, (see e.g. \citealt{2022MNRAS.509.1447M} and references within). 
%So, considering BIC and AIC for each model seems to be the way to go. 
As pointed out in various works (e.g. \citealt{2017AJ....154....4P,2022MNRAS.509.1447M}), the difference between the BIC and AIC values of two models, $\eta= \Delta {\rm BIC}$ and $\xi= \Delta {\rm AIC}$, are good statistical indicators. In particular, $\eta>10$ and a $\xi>10$ indicate that the evidence favouring the lower BIC model against the other is very strong. 

\begin{figure}[h]
\centering
\includegraphics[width=\hsize]{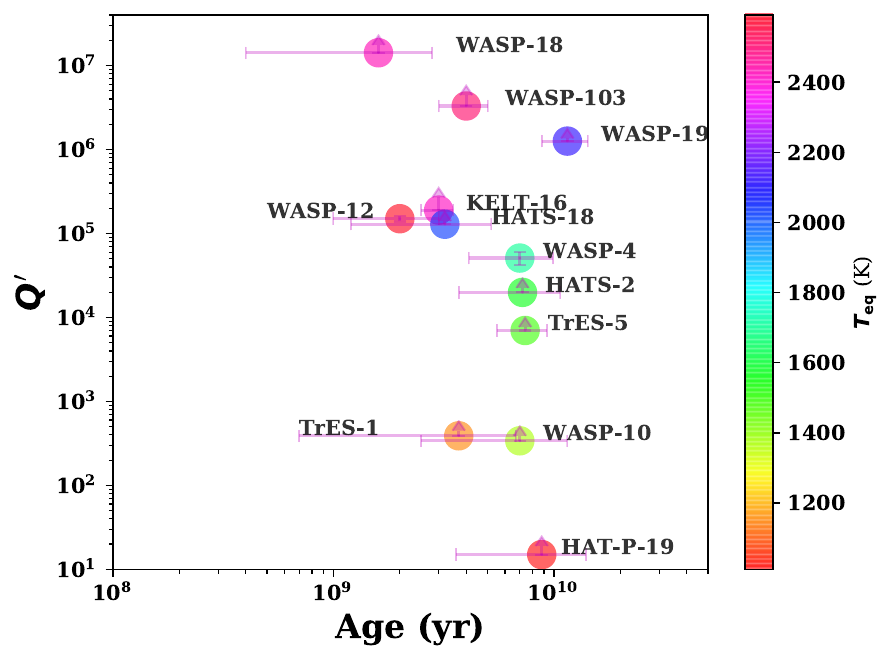}
\caption{Stellar tidal quality factor as a function of stellar age for HATS-2 and an ensemble of host stars. These stars were selected considering systems where tidal decay has been observed or is a possible explanation for the observed TTVs (\citealt{hagey}; \citealt{rosario}; \citealt{wong}; \citealt{2022MNRAS.509.1447M}; \citealt{barros}).  }
\label{fig:q}
\end{figure}

First, we focus on the comparison between the linear and the quadratic ephemeris models. Without the ExoClock data, the quadratic model is favoured by a $\eta=9.6$ and a $\xi=10.7$, so we are at the limit of the statistical significance. Adding the ExoClock data, the quadratic model is favoured by a $\eta=15.1$ and a $\xi=16.4$. So, the tidal-decay scenario seems to be highly preferred with respect to a simple linear-ephemeris scenario and this makes HATS-2 a good candidate in the context of the search for secular timing variations in hot Jupiters (\citealt{hagey}). Focusing now on the second scenario results, by using the values listed in Table \ref{table:coeff} and the following formula \citep{2017AJ....154....4P}:
\begin{equation}
\frac{1}{2}\frac{{\rm d} P}{{\rm d}E}E^2= -\frac{27\pi}{4\,Q_{\star}^{\prime}}\frac{M_{\rm p}}{M_{\star}} \left(\frac{R_{\star}}{a}\right)^5 P E^2 \, ,
\label{eqn:decay}
\end{equation}
we derived a $\frac{dP}{dE}=-(5.48 \pm 2.48)\times 10^{-10}$ days per orbital cycle, and therefore the period derivative is $\dot{P}=\frac{1}{P} \frac{dP}{dE}=-15.7 \, \pm \, 3.5$ ms yr$^{-1}$,
consistent with an orbital period that shrink to zero in a time within $\frac{P}{dP/dt}=\frac{P^2}{dP/dE}= 7.5\, \pm \, 1.7$ Myr. This is reasonable for a planet that orbits around its host star with a semi-major axis $\simeq 2$ times the Roche radius and so close to tidal disruption. Using Eq.~(\ref{eqn:decay}), we also derived a limit to the modified stellar tidal dissipation quality factor of $Q_{\star}^{\prime} > (1.99 \pm 0.26)\times 10^4$; this result is based on the 95 per cent confidence lower limits on $\dot{P}$, while the uncertainties come from propagating the errors in $M_{\rm p}/M_{\star}$ and $R_{\star}/a$. 
This result is consistent with theoretical works such that by \cite{ahuir}. In Fig. \ref{fig:q}, we plot the obtained limit as a function of stellar age and planetary equilibrium temperature, together with the values obtained for other systems hosting hot Jupiters that can be considered good candidates to study orbital decay. These are: KELT-16 \citep{2022MNRAS.509.1447M}, HATS-18 \citep{2022MNRAS.515.3212S}, WASP 18 and WASP 19 \citep{rosario}, WASP-12 \citep{wong}, WASP-4 \citep{turner}, WASP-103 \citep{barros}, and TrES-1, TrES-5 and HAT-P-19 \citep{hagey}. The stellar ages were taken from \cite{bonomo}. In \cite{hagey} only the orbital period derivates were presented, so we had to compute the $Q_{\star}^{\prime}$ limits via Eq.~(\ref{eqn:decay}); the planetary and stellar parameters have been taken from the literature (i.e. \citealt{tres1,tres5,wasp10,hatp19}). The results show a correlation between the derived tidal quality factor and the planetary equilibrium temperature, in agreement with Eq.~(\ref{eqn:decay}). In this context, the value derived for HATS-2 occupies the right place in the diagram of the tidal quality factor versus stellar age and equilibrium temperature. The relation between the stellar tidal quality factor and the stellar age is not clear and further observations may give new insights.

Investigating Table~\ref{table:chired}, it is possible to observe that also a sinusoidal model can explain the data well. In fact, in both scenarios, the relative $\eta$ and $\xi$ are always smaller than 5. The sinusoidal model can be connected to the presence of apsidal precession induced by the tidal interactions between the planet and the host star (\citealt{wolf,gimenez}):
\begin{equation}
{\rm O-C} \simeq -\frac{eP_{\rm a}}{\pi}\,\cos{\omega}=-\frac{eP_{\rm a}}{\pi}\,\cos{\left( \omega_0+\frac{d\omega}{dE}E\right)} \, ,
\label{eqn:apsoc}
\end{equation}
\begin{equation}
\frac{d\omega}{dE}\simeq15\pi\,k_{2,p}\frac{M_{\star}}{M_{\rm p}}\left(\frac{R_{\rm p}}{a}\right)^5 \, ,
\label{eqn:apsfinal}
\end{equation}
where $P_{\rm a}$ indicates the sidereal period, $\omega$ the argument of the periastron, $k_{2,p}$ is the planet's Love number and $d\omega/dE$ the precession rate. Plugging the values listed in Table \ref{table:coeff} into Eq.~(\ref{eqn:apsfinal}), we obtained a value of $k_{2,p}$ greater than 1.5 that has no physical meaning. The other physical explanation is the presence of a third body in the system (\citealt{agol}). However, modelling TTVs induced by a third body is a highly degenerate problem and the quality of our data and the temporal coverage are not good enough to face it. In any case, future and systematic recordings of new mid-transit times will certainly give useful insights to entangle the problem.
Fig.~\ref{fig:hats2eff} tells us that it should be possible to discriminate between the quadratic and the sinusoidal model by exploiting new transit measures in the next two years.

%-----------------------------
\section{Variation in the planetary radius with wavelength}
\label{sec:atm1}
%-----------------------------
Due to their proximity to their parent star, hot Jupiters are strongly irradiated and their spectra are expected to show several absorption features at optical wavelengths. Some of the features that are usually detected in the optical bands include sodium ($\sim 590$\,nm), potassium ($\sim 770$\,nm) and water vapour ($\sim 950$\,nm). In several cases, Rayleigh scattering at bluer wavelengths has been found as well as the presence of strong absorbers, such as gaseous titanium oxide (TiO) between 450 and 800\,nm.

The transit depth is a parameter that depends on the wavelength and, therefore, on the possible absorption of the chemical species present in the exoplanetary atmosphere. Multi-colour photometry can be a useful tool for investigating how $k$ changes with wavelength, probing in this way the chemical composition at the terminator of planetary atmospheres during transit events. 
\emph{Transmission photometry} has, in general, a very low spectral resolution but is suitable for ground-based telescopes with smaller apertures and exoplanets orbiting faint stars.
This type of study can be, therefore, useful to get information about the absorption behaviour of the exoplanetary atmosphere and guide more precise investigation via \emph{transmission spectroscopy}, being the latter a most effective technique.

%An alternative approach for probing planetary atmospheres is that of  Moreover, photometric observations are much less affected by telluric contamination than spectroscopic ones. The working principle is the same as transmission spectroscopy but does not require the use of a spectrometer. Instead, we can use multi-color photometry, for which is possible to average the $k$ derived from the light curves taken with the same filters and plot them against the wavelength of the used filter. In this way we obtain a transmission spectrum with a lower resolution with respect to the transmission spectroscopy but, if the quality of our data is good, is still possible to get information about the global absorption behavior of the exoplanetary atmosphere by comparing it with theoretical models. 

\begin{figure}
\centering
\includegraphics[width=\hsize]{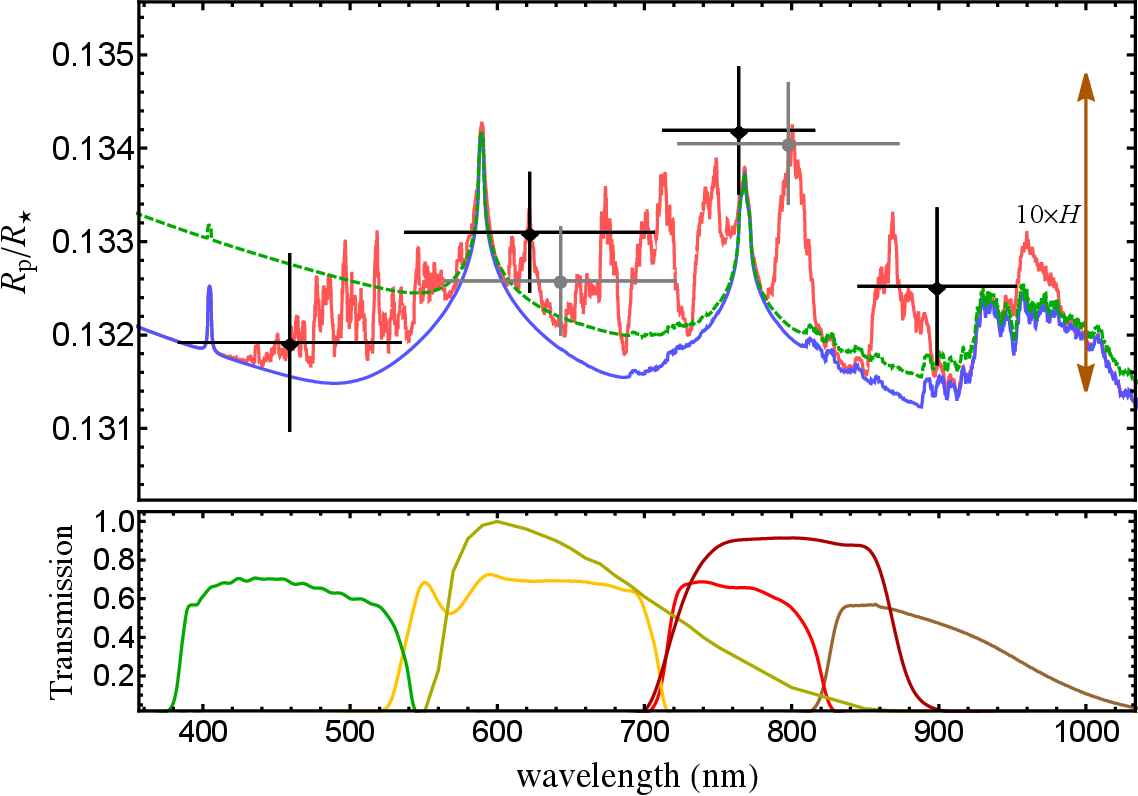}
\caption{{\it Top panel}: Variation in the planetary radius, in terms of the planet-to-star radius ratio, with wavelength. The points are from the ground-based transit observations presented in this work (black from the MPG\,2.2\,m telescope and grey from the Danish 1.54\,m telescope). The vertical bars represent the errors in the measurements and the horizontal bars show the FWHM transmission of the passbands used. The observational points are compared with three synthetic spectra from \citet{2010ApJ...709.1396F} (Red: equilibrium-chemistry; Blue: no TiO, VO opacity; Green: enhanced Rayleigh scattering). The same offset is applied to all four models to provide the best fit to the measurements. The size of 10 atmospheric pressure scale heights ($10 \times H$) is shown on the right of the plot.
{\it Bottom panel}: Transmission curves for the Bessell $I$ and $R$ filters and the total efficiencies of the GROND filters.}
\label{fig:spectrum}
\end{figure}

Using the light curves of the transits of HATS-2\,b, which we obtained through observations at different passbands, we attempted to get a low-resolution transmission spectrum of HATS-2\,b. Following a general approach, we run PRISM/GEMC for each of the ground-based light curves again to calculate the ratio of the radii in each passband, being the other photometric and starspots parameters fixed to the best-fitting values listed in Tables \ref{tab:fit} and \ref{tab:fitsp}. This yielded a set of $k$ values which are directly comparable and whose error bars exclude common sources of uncertainty. In particular, for fixing the values of the LD coefficients, we calculated the weighted means of the values, based on their colour. As a last step, we averaged the ratio of the radii values taken with the same filters. The weighted mean was performed using the scatter in the light curves to set the size of the errorbar. In this way, we can be sure that the uncertainties will not be underestimated. We excluded the TESS light curves because the long-pass filter of TESS is just too wide ($>500$\,nm) for our purposes. The results are: 
\begin{itemize}
\item[] $k_{g'}= 0.13192\pm 0.00094$; 
\item[] $k_{r'}= 0.13310\pm 0.00063$; 
\item[] $k_{R}=0.13258 \pm 0.00057$;
%(il k ottenuto mediando r' con R è 0.13345 +/- 0.00025), 
\item[] $k_{i'}= 0.13419 \pm 0.00067$; 
\item[] $k_{I}= 0.13405 \pm 0.00064$;
%(il k ottenuto mediando i' con I è 0.134185 +/- 0.00026), 
\item[] $k_{z'}= 0.13252 \pm 0.00083$. 
\end{itemize}
%Having transits recorded with the GROND camera (Sloan filters) and with the DFOSC instrument, equipped with Bessell $R$ and $I$ filters, we have a coverage of $\simeq 600$ nm. 
These values are shown in Fig.~\ref{fig:spectrum} and are compared with three 1D model atmospheres, which were obtained by \citet{2010ApJ...709.1396F}. 

In particular, the red line represents an equilibrium-chemistry spectrum, which was calculated for Jupiter’s gravity ($25$\,m\,s$^{-2}$) with a base radius of $1.25\,R_{\rm Jup}$ at the 10 bar level and at 1500\,K. This model is dominated by H$_2$/He Rayleigh scattering in the blue and pressure-broadened neutral atomic lines of sodium and potassium at 770\,nm, and the opacity of TiO and VO molecules were included.
The blue line represents a model similar to the previous one but the opacity TiO and VO were excluded. Another spectrum was computed from the latter by changing the intensity of Rayleigh scattering, which physically corresponds to considering different compositions and thicknesses of the external atmosphere layers. In particular, the dashed dark-green line was obtained by increasing the Rayleigh scattering by a factor of 10.

A variation in the planetary radius was found between the $g^{\prime}$ and $i^{\prime}$ bands.
% or between the $z^{\prime}$ and $i^{\prime}$ bands. In both cases, 
The variation is roughly 6.5 pressure scale heights\footnote{The pressure scale height of a planetary atmosphere is defined as $H = k_{\rm B} T_{\rm eq} / \mu_{\rm m} g_{\rm p}$, where $k_{\rm B}$ is the Boltzmann constant and $\mu_{\rm m}$ is the mean molecular weight, for which the value of 2.3\,amu is usually adopted for giant planets \citep{Lecavelier2008}.} but its significance is below the $2\,\sigma$ level, which preserve us to claim any absorption detection.

%\begin{figure}[h]
%	\centering
%	%\hspace*{-2cm}
%	\includegraphics[width=1.1\linewidth]{eqchem.pdf}
%	\caption{Variation of the planetary radius, in terms of the planet/star radius ratio, with wavelength. The points are from the ground-based transit observations presented in this work, and their color scheme is explained in the plot legend. The vertical bars represent the errors in the measurements and the horizontal bars show the FWHM transmission of the passbands used. The observational points are compared with a synthetic spectrum from \cite{fortney2010} (pink line), which was generated considering an equilibrium chemistry scenario between optical absorbers as TiO, and alkani metals as K and Na.}
%	\label{fig:eqchem}
%\end{figure}
%
%\begin{figure}[h]
%	\centering
%	%\hspace*{-2cm}
%	\includegraphics[width=1.1\linewidth]{noTiO.pdf}
%	\caption{Variation of the planetary radius, in terms of the planet/star radius ratio, with wavelength. The points are from the ground-based transit observations presented in this work, and their color scheme is explained in the plot legend. The vertical bars represent the errors in the measurements and the horizontal bars show the FWHM transmission of the passbands used. The observational points are compared with synthetic spectra from \cite{fortney2010}, which were generated considering only the presence of K and Na as principal absorbers and the effects of Rayleigh Scattering at different levels of intensity (see the plot legend for further details).}
%	\label{fig:notio}
%\end{figure}

%-----------------------------
\section{Flux ratio of HATS-2 b from TESS phase curve}
\label{sec:atm2}

\begin{figure}[h]
\centering
\includegraphics[width=\hsize]{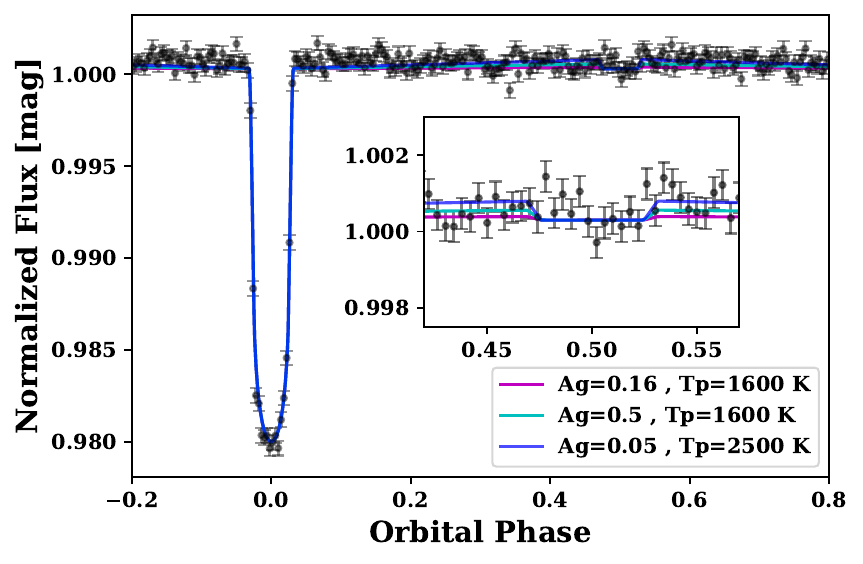}
\caption{HATS-2b phase curve obtained by phase-binning the data from Sectors 10, 36 and 63 of NASA's TESS mission plotted against theoretical curves computed thanks to the {\texttt{batman}} Python package.  }
\label{fig:eclipse}
\end{figure}

We investigated the possibility to measure the occultation depth, in other words the planet-to-star flux ratio $F_{\rm p}/F_{\star}$, in the continuous time series data of TESS (see Fig.~\ref{fig:tess}). For this purpose, we phased the data from every sector considering the best-fitting linear ephemeris found in the previous sections, Eq.\,(\ref{eqn:eff2}), and binned them using a weighted sigma-clipping mean, which means that the data are iteratively kept or rejected in terms of standard deviation. The resulting phase curve is presented in Fig. \ref{fig:eclipse}, together with different theoretical phase curves generated thanks to the {\texttt{batman}} Python package \citep{batman}. To simulate the occultation curves, {\texttt{batman}} needs two free parameters, the occultation depth and the occultation mid-time, and a series of fixed parameters as the mid-transit time, the ratio of the radii, the scaled stellar radius, the period, the eccentricity, and the inclination of the orbit. In turn, the occultation depth at visible wavelengths (such as the one at which the TESS bandpass is centred, $\simeq$ 786 nm) can be connected to the planetary day-side temperature $T_{\rm p}$ and geometric albedo $A_{\rm g}$ \citep{kovacs}:
\begin{equation}
\delta_{\rm occ} \simeq \delta_{\rm refl} + \delta_{\rm th} \simeq A_{\rm g} \left( \frac{R_{\rm p}}{a} \right) ^2 + k^2 \frac{B_{\nu}(T_{\rm p})}{B_{\nu}(T_{\rm eff})},
\label{eqn:occdepth}
\end{equation}
where $B_{\nu}$ is the Planck function. If we use Eq.~(\ref{eqn:occdepth}) together with the values listed in Table \ref{table:jktabsdim} we can generate different theoretical light curves considering different scenarios and corresponding values for $A_{\rm g}$ and $T_{\rm p}$. In particular, we considered three scenarios: ($i$) $A_{\rm g}=0.16$ and $T_{\rm p}=1600$\,K, ($ii$) $A_{\rm g}=0.5$ and $T_{\rm p}=1600$\,K, and ($iii$) $A_{\rm g}=0.05$ and $T_{\rm p}=2500$\,K. The first scenario should be that closer to reality if we consider typical values of $A_{\rm g}$ found in the literature for hot Jupiters and a conservative atmospherical circulation. The second and third scenarios, instead, represent two opposite extremes. The second represents the situation where most of the light that we receive from the planet in the TESS bandpass is just reflected light; the third, instead, represents the case where the day-side temperature greatly exceeds the equilibrium temperature, for example, due to the presence of absorbers or tidal locking. In any case, none of the generated theoretical curves seems to be close to the phased TESS light curve for which a non-detection of the occultation is the most compatible scenario. Moreover, the uncertainty on the mean flux (unbinned) is close to 1 ppt (1000\,ppm), which makes any tentative of measuring the occultation depth extremely difficult. In the end, as expected due to the faintness of the parent star, we did not report any detection of a secondary eclipse in the HATS-2 TESS photometry. This work and TESS photometry place an upper limit on the occultation depth of HATS-2b, $\delta_{\rm occ}< 400$\,ppm.
%-----------------------------

\section{Summary and conclusions}
\label{sec:sum}
In this work, we studied the physical and orbital properties of the hot Jupiter HATS-2\,b, a giant transiting exoplanet with $P_{\rm orb} \approx 1.35$\,days, and therefore, an interesting target to investigate in relation to a possible orbital-decay process. We reported the photometric monitoring
of 13 new transit events of HATS-2\,b, which were observed with two medium-class telescopes through six different optical passbands (see Figs.~\ref{fig:dan} and \ref{fig:grond}).
The transits were observed using the defocusing technique, achieving a photometric precision of 0.6\,mmag per observation in the best case. Two transits were simultaneously observed in four optical bands thanks to the GROND multi-band camera. In total, we collected 19 new light curves. We also considered the data from two other transits observed with GROND \citep{2013A&A...558A..55M} and four published and two new light curves (see Fig.~\ref{fig:exoclock}) that were recorded by amateur astronomers within the ExoClock project. We also considered the photometric data collected by TESS and analysed 48 complete transits recorded by this space telescope (see Fig.~\ref{fig:tess}). Most of the light curves presented features that break the classical symmetry of transit events. Such features are attributable to starspot complexes that are revealed by the planet HATS-2\,b when it transits in front of its parent star, and thus making a temperature scan of a chord of the stellar photosphere. The main results that we obtained are as follows.
\begin{itemize}
\item
Thanks to the TESS light curves and the new ground-based light curves, we reviewed the main physical and orbital parameters of the HATS-2 planetary system. Our results are shown in Table~\ref{table:jktabsdim} and are in very good agreement with those obtained by \citet{2013A&A...558A..55M}.  \\ [-6pt]
\item
Almost all of the light curves were analysed with PRISM+GEMC, a code able to simultaneously fit a transit light curve containing one or more starspot features. This modelling of the light curves allowed us to determine the position of the starspots on the stellar disc as well as their size and contrast (see Table~\ref{tab:fitsp}). Having monitored several transits in different filters, we verified that the variations in the starspot contrast with wavelength were in agreement with the expectations from theoretical models (see Fig.~\ref{fig:contrastsub}). Based on theoretical assumptions, we also estimated the temperature of each starspot (see Table~\ref{tab:fitsp}).  \\ [-6pt]
\item
We detected starspot features on the light curves of consecutive transits of HATS-2\,b both from the ground and from space. In particular, we identified
ten and one consecutive starspot-crossing events from the space and ground observations, respectively. All these transits are separated by less than five days, and the latitudes and angular size differences of the starspots fall within $1\sigma$. They also have similar amplitudes, contrast and duration. We concluded that for each of these 11 consecutive transits, the same starspot was occulted by the planet after having rotated a bit on the surface of the parent star. Based on this sensible hypothesis, we easily determined the stellar rotational period, $P_{\rm rot}=22.46 \pm 5.20$\,days, and the sly-projected spin-orbit obliquity, $\lambda=2^{\circ}.72 \pm 17^{\circ}.84$. We also 
placed a weak constraint on the true orbital obliquity $\psi=38^{\circ}.49 \pm 27^{\circ}.13$.  \\ [-6pt]
\item
We estimated the mid-transit time for each of the transit events of HATS-2\,b that we presented plus others from the ExoClock archive, obtaining a list of 27 epochs that covers a time baseline extending over more than ten years (see Table~\ref{table:eff}). We used these timings to update the ephemeris of the orbital period and to search for possible variations in the mid-transit times. We found that models that predict both orbital decay and apsidal precession fit the data better than the linear-ephemeris scenario. The observations of many other transits of HATS-2\,b and new radial-velocity measurements are needed to firmly confirm our indication that its orbit is decaying or is affected by a third body in the system.  \\ [-6pt]
\item
Thanks to the multi-band photometric observations of HATS-2\,b transits, we were able to obtain a low-resolution optical transmission spectrum of the planet. We found a variation in the planetary radius between the $g^{\prime}$ and $i^{\prime}$ bands. This variation is roughly 6.5 pressure scale heights but its significance is below the $2\,\sigma$ level. Therefore, we cannot claim the presence of strong optical absorbers (either Na and K, or TiO) in the atmosphere of HATS-2\,b. More accurate data are needed for this purpose. \\ [-6pt]
\item 
We analysed the continuous HATS-2 time series data of TESS to measure the occultation depth,
that is the planet-to-star flux ratio. We obtained a phase curve of HATS-2\,b by phase-binning the data from three sectors, and then we compared it against different theoretical models testing various values for the planetary day-side temperature and geometric albedo. Our analysis did not lead to any conclusive results since the occultation is not seen in the TESS data and none of the theoretical curves is able to explain the phased TESS light curve.

\end{itemize}
%
%-----------------------------
\begin{acknowledgements}
This work is based on data collected with the MPG\,2.2\,m telescope and within the MiNDSTEp program with the Danish 1.54\,m telescope. Both telescopes are located at the ESO Observatory in La Silla (Chile). The GROND camera was built by the high-energy group of MPE in collaboration with the LSW Tautenburg and ESO, and is operated as a PI instrument at the MPG\,2.2\,m telescope.
This work includes data collected with the TESS mission, obtained from the MAST data archive at the Space Telescope Science Institute (STScI). Funding for the TESS mission is provided by the NASA Explorer Program. STScI is operated by the Association of Universities for Research in Astronomy, Inc., under NASA contract NAS 5–26555.
We acknowledge the use of public TESS data from pipelines at the TESS Science Office and at the TESS Science Processing Operations Center. Resources supporting this work were provided by the NASA High-End Computing (HEC) Program through the NASA Advanced Supercomputing (NAS) Division at Ames Research Center for the production of the SPOC data products.
We acknowledge funding from the Novo Nordisk Foundation Interdisciplinary Synergy Program grant no. NNF19OC0057374 and from the European Union H2020-MSCA-ITN-2019 under grant No. 860470 (CHAMELEON).
Support for this project is provided by ANID's Millennium Science Initiative through grant ICN12\textunderscore 009, awarded to the Millennium Institute of Astrophysics (MAS), and by ANID's Basal project FB210003.
L.\,M. acknowledges support from the MIUR-PRIN project No. 2022J4H55R.
T.C.H. acknowledges funding from the European Union’s Horizon 2020 research and innovation programme under grant agreement No.\,871149. The Horizon 2020 programme is supported by the Europlanet 2024 RI provides free access to the world's largest collection of planetary simulation and analysis facilities, data services and tools, a ground-based observational network and programme of community support activities.
%
%Part of this work was supported by the German
%\emph{Deut\-sche For\-schungs\-ge\-mein\-schaft, DFG\/} %project number Ts~17/2--1.
%
The following internet-based resources were used in research for this paper: the ESO Digitized Sky Survey; the NASA Astrophysics Data System; the SIMBAD database operated at CDS, Strasbourg, France; and the ar$\chi$iv scientific paper preprint service operated by the Cornell University.
\end{acknowledgements}

% WARNING
%-------------------------------------------------------------------
% Please note that we have included the references to the file aa.dem in
% order to compile it, but we ask you to:
%
% - use BibTeX with the regular commands:
%   \bibliographystyle{aa} % style aa.bst
%   \bibliography{Yourfile} % your references Yourfile.bib
%
% - join the .bib files when you upload your source files
%-------------------------------------------------------------------
\bibliographystyle{aa}
\bibliography{bib}

\begin{appendix}
\section{PRISM/GEMC and JKTEBOP fits of TESS data}

\FloatBarrier

\begin{figure}[h]
\centering
\includegraphics[width=\hsize]{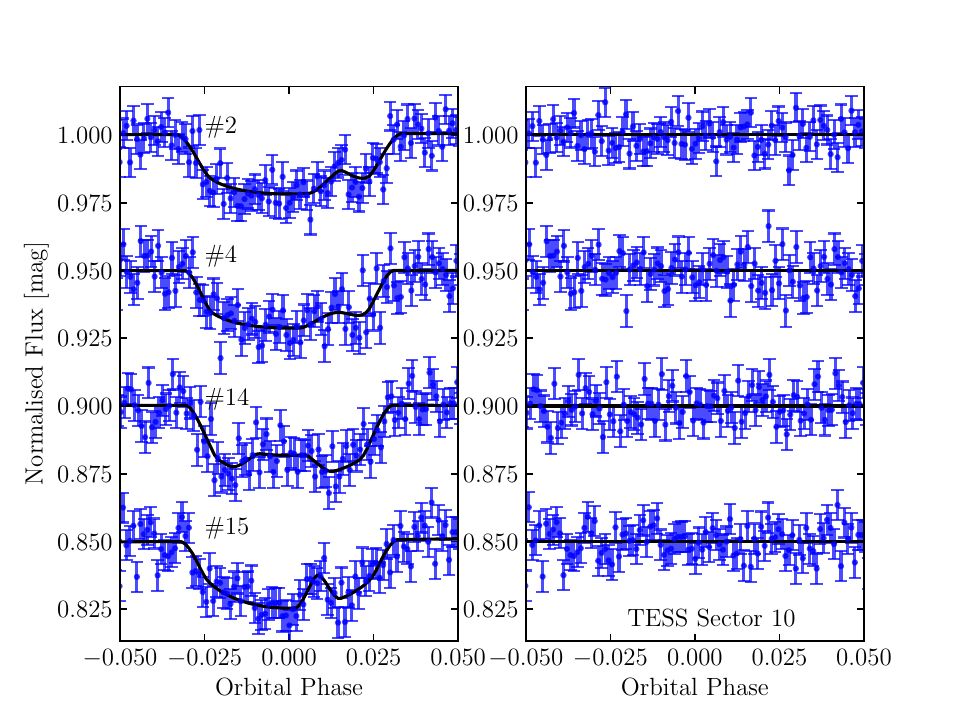}
\caption{{\it Left panel}: Light curves of HATS-2 obtained from the Sector 10 of NASA’s TESS mission data and used in the analysis of the physical and starspot parameters of the system. They are plotted against the orbital phase and are compared to PRISM/GEMC best-fitting models. The labels indicate the observational ID (see Fig.~\ref{fig:tess}). {\it Right panel}: Residuals of the fits represented with the same notation used in the left panel.}
    \label{fig:tess1}
\end{figure}

\begin{figure}[h]
\centering
\includegraphics[width=\hsize]{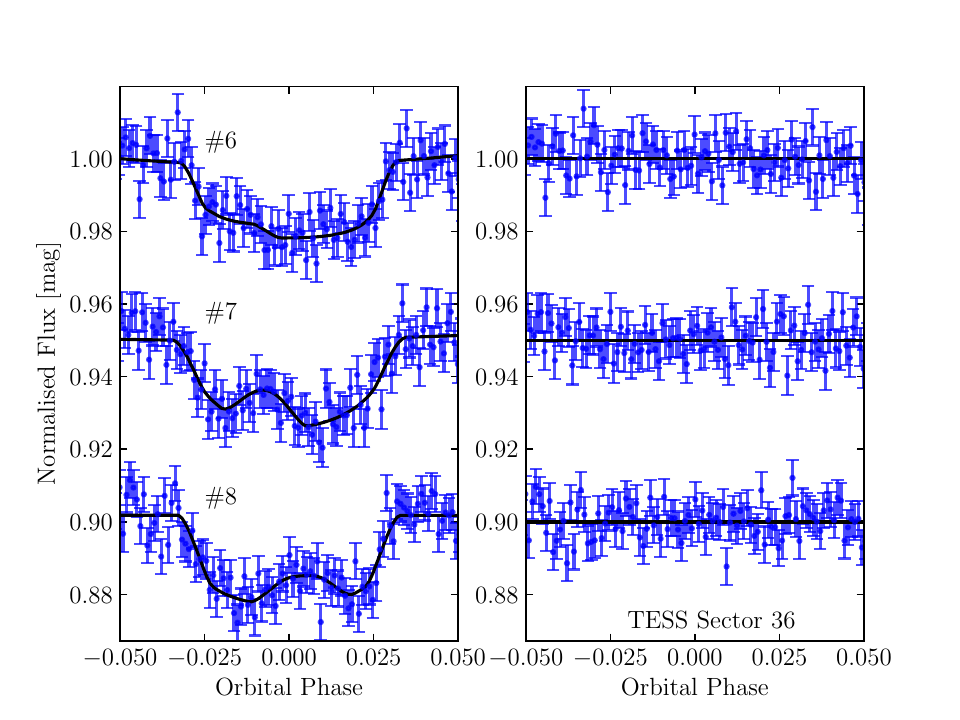}
\caption{
{\it Left panel}: Light curves of HATS-2 obtained from the Sector 36 of NASA’s TESS mission data, and used in the analysis of the physical and starspot parameters of the system. 
In particular, the image refers only to the transits labelled $\#6$, $\#7$, and $\#8$ (see Fig. \ref{fig:tess} and Fig. \ref{fig:sp}). They are plotted against the orbital phase and are compared to PRISM/GEMC best-fitting models. {\it Right panel}: Residuals of the fits represented with the same notation used in the left panel.
}
    \label{fig:tess3}
\end{figure}

\begin{figure}[h]
    \centering
    \includegraphics[width=\hsize]{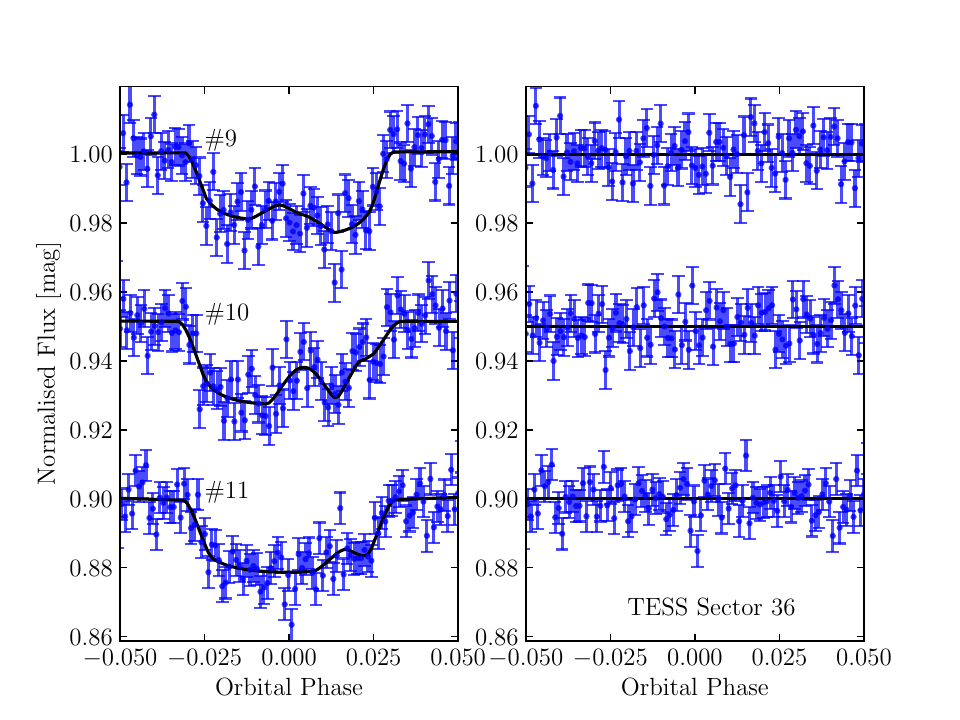}
    \caption{{\it Left panel}: Remaining transit light curves containing spot-crossing events from Sector 36 of NASA's TESS mission data that were not represented in Fig. \ref{fig:tess3}. They are plotted against the orbital phase and are compared to new PRISM/GEMC best-fitting models. The labels indicate the observational ID. {\it Right panel}: Residuals of the fits represented with the same notation used in the left panel.}
    \label{fig:tess4}
\end{figure}

\begin{figure}[h]
\centering
\includegraphics[width=\hsize]{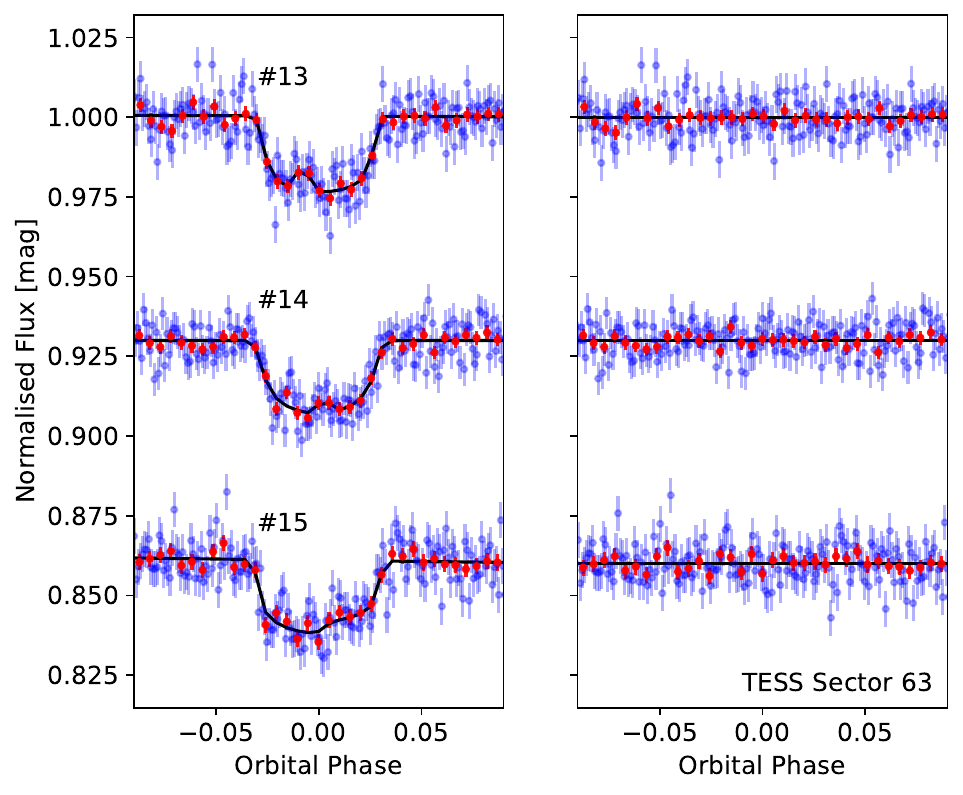}
\caption{
{\it Left panel}: The light curves of HATS-2 obtained from Sector 63 of NASA’s TESS mission data, and used in the analysis of the physical and starspot parameters of the system. In blue are represented the unbinned data, while in red the 10 min cadence binned data.
In particular, the image refers only to the transits labelled $\#13$, $\#14$, and $\#15$ (see Fig. \ref{fig:tess}). They are plotted against the orbital phase and are compared to PRISM/GEMC best-fitting models. {\it Right panel}: Residuals of the fits represented with the same notation used in the left panel.
}
    \label{fig:tess6}
\end{figure}

\FloatBarrier

\begin{figure}
    \centering
    \hspace*{-0.7cm}
    \includegraphics[width=\hsize]{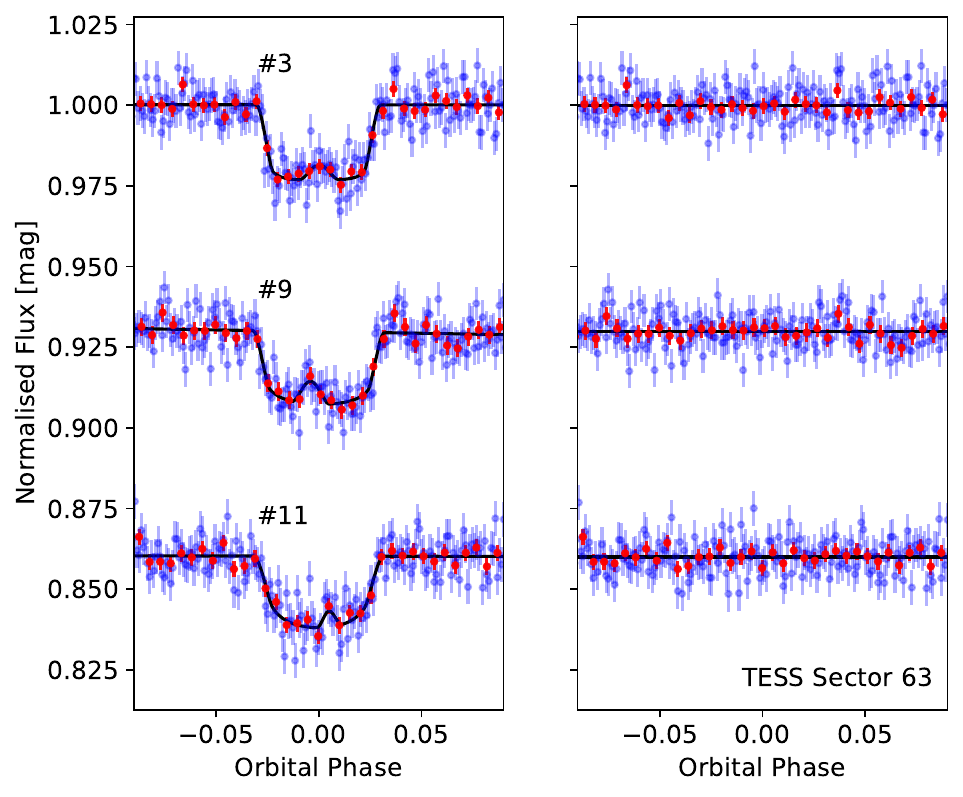}
    \caption{{\it Left panel}: Remaining transit light curves containing spot-crossing events from Sector 63 of NASA's TESS mission data that were not represented in Fig. \ref{fig:tess6}. In blue are represented the unbinned data, while in red the 10 min cadence binned data. They are plotted against the orbital phase and are compared to new PRISM/GEMC best-fitting models. The labels indicate the observational ID. {\it Right panel}: Residuals of the fits represented with the same notation used in the left panel.}
    \label{fig:tess7}
\end{figure}

\begin{figure}
    \centering
    \hspace*{-1.8cm}
    \includegraphics[width=1.3\columnwidth]{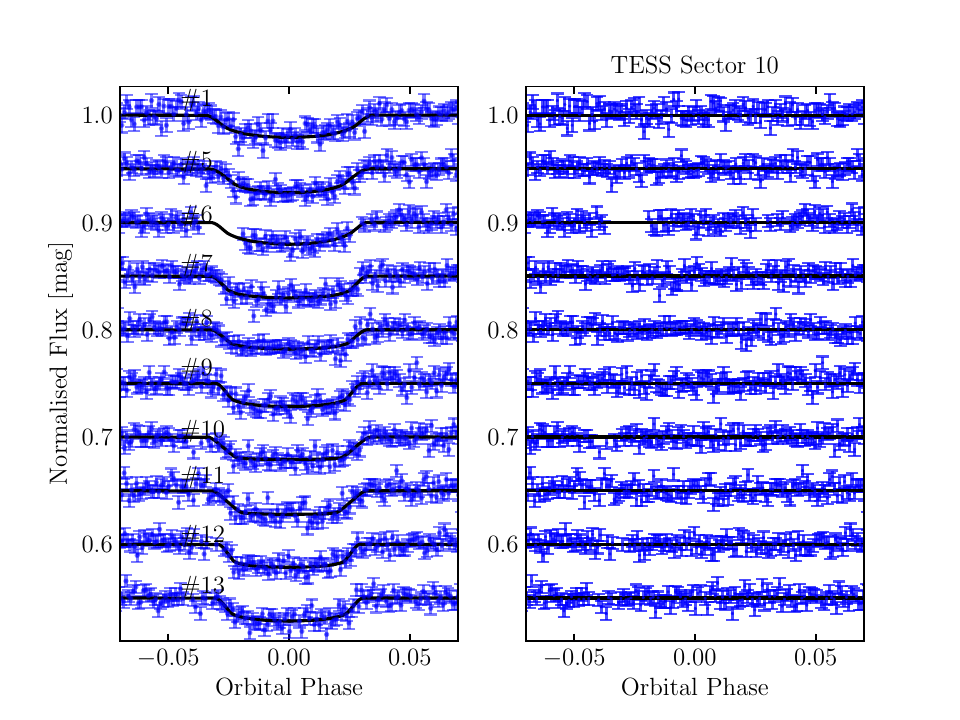}
    \caption{{\it Left panel}: Light curves of eight transits of HATS-2b observed with NASA’s TESS space telescope in Sector 10 of its primary mission. The labels indicate the observation ID (see Fig. \ref{fig:tess}). They are plotted against the orbital phase and are compared to the best-fitting JKTEBOP models. {\it Right panel}: Residuals of each fit.}
    \label{fig:tess2}
\end{figure}

\begin{figure}
    \centering
    \hspace*{-0.5cm}
    \includegraphics[width=1.3\columnwidth]{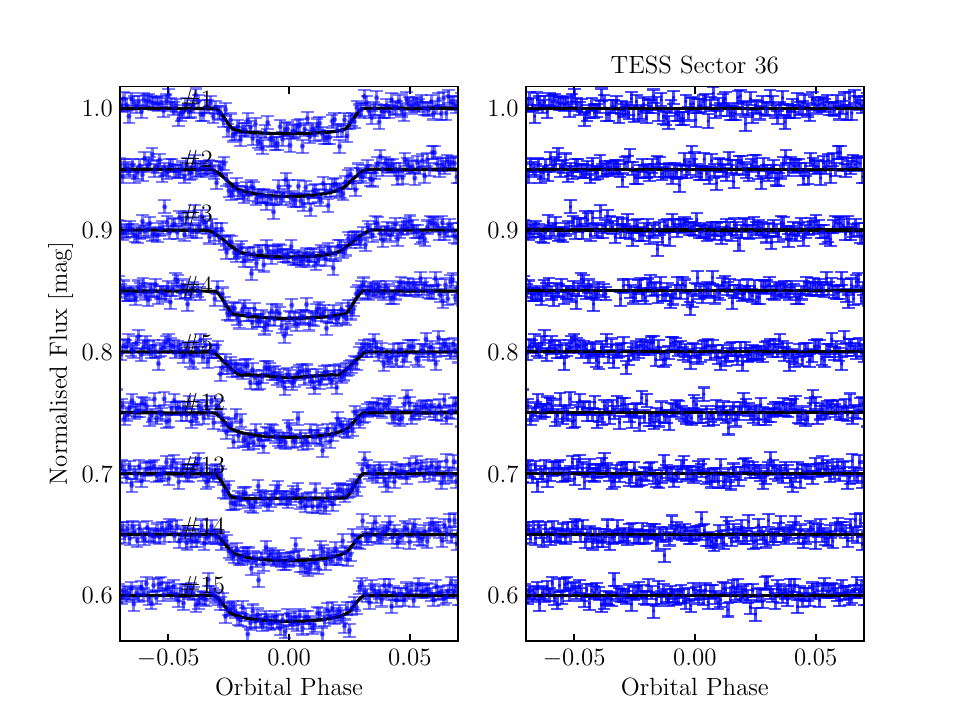}
    \caption{{\it Left panel}: Light curves of eight transits of HATS-2b observed with NASA’s TESS space telescope in Sector 36 of its primary mission. The labels indicate the observation ID (see Fig. \ref{fig:tess}). They are plotted against the orbital phase and are compared to the best-fitting JKTEBOP models. {\it Right panel}: Residuals of each fit.}
    \label{fig:tess5}
\end{figure}

\FloatBarrier

\begin{figure*}
    \centering
    \includegraphics[width=\hsize]{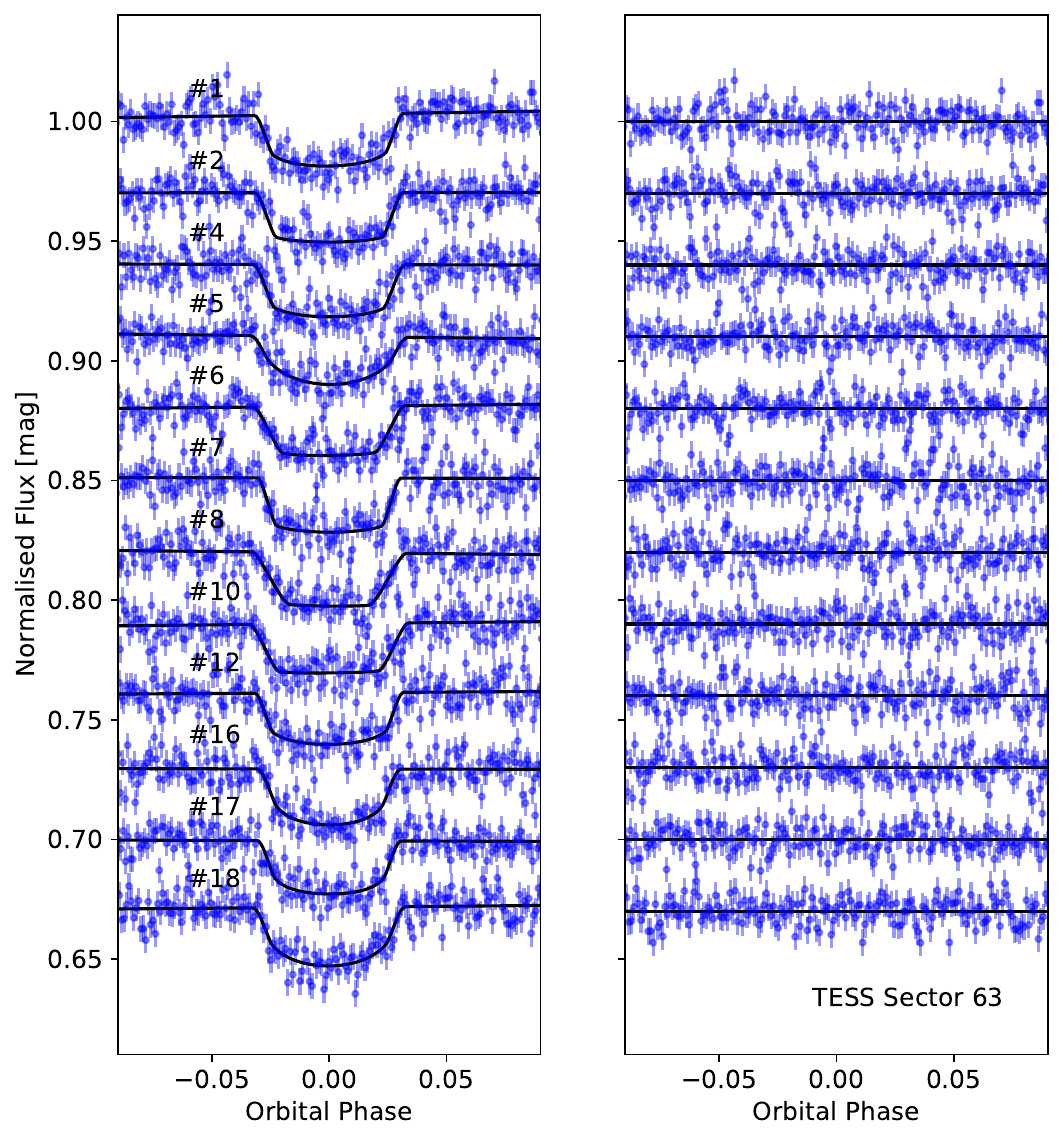}
    \caption{{\it Left panel}: Light curves of 12 transits of HATS-2b observed with NASA’s TESS space telescope in Sector 63 of its primary mission. The labels indicate the observation ID (see Fig. \ref{fig:tess}). They are plotted against the orbital phase and are compared to the best-fitting PRISM+GEMC models. {\it Right panel}: Residuals of each fit.}
    \label{fig:tess8}
\end{figure*}

\FloatBarrier

\section{Mathematical derivation of Section 4.3 equations}

Using spherical trigonometry, the relation that connects the longitude $\phi$ and latitude $\theta$ to Cartesian coordinates for a point on a sphere is:

\begin{equation}
    \begin{cases}
      x=R_{\star} \cos \theta \cos \phi\\
      y=R_{\star} \cos \theta \sin \phi \\
      z=R_{\star} \sin \theta
    \end{cases}\,.
\end{equation}

\begin{figure}[h]
    \centering
    \includegraphics[width=\hsize]{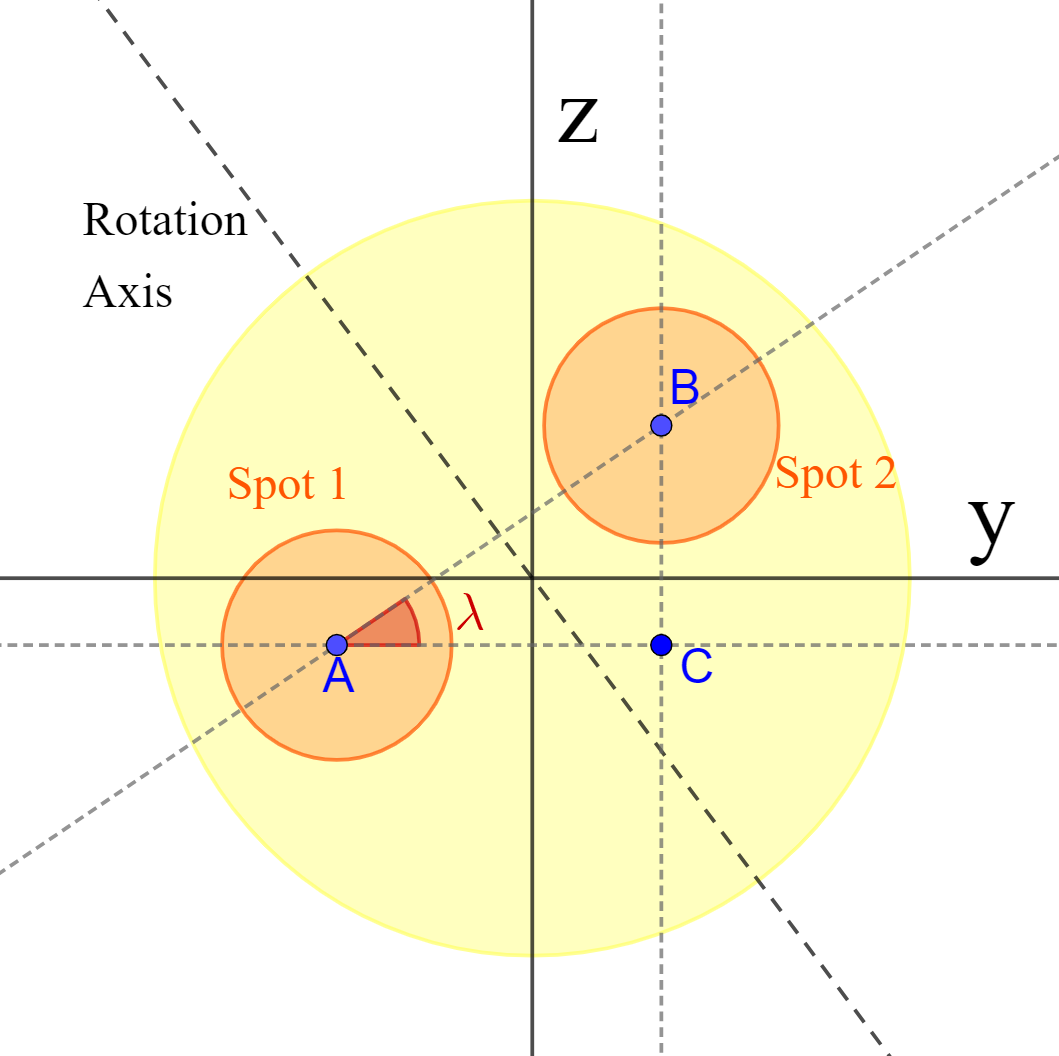}
    \caption{Geometry of the problem. The rotation axis differs from the normal to the sky-projected plane, so two consecutive spots are observed at different latitudes and longitudes.}
    \label{fig:geo}
\end{figure}

Referring to Fig. \ref{fig:geo}, the sky-projected obliquity is given by:

\begin{equation}
\lambda= \arctan{\frac{\rm{BC}}{\rm{AC}}}=\arctan{\frac{z_2 - z_1}{y_2 - y_1}}.
\end{equation}

To obtain the rotational period a proportion must be considered: $P_{\rm rot}\, : \,2\pi R_{\star} = \Delta t \,: \, \rm{AB}$. Using Pythagorean theorem is possible to derive the Equation in Section 4.3.

\FloatBarrier

\section{GLS fits of TESS photometry}

\begin{figure}[h]
    \centering
    \includegraphics[width=\hsize]{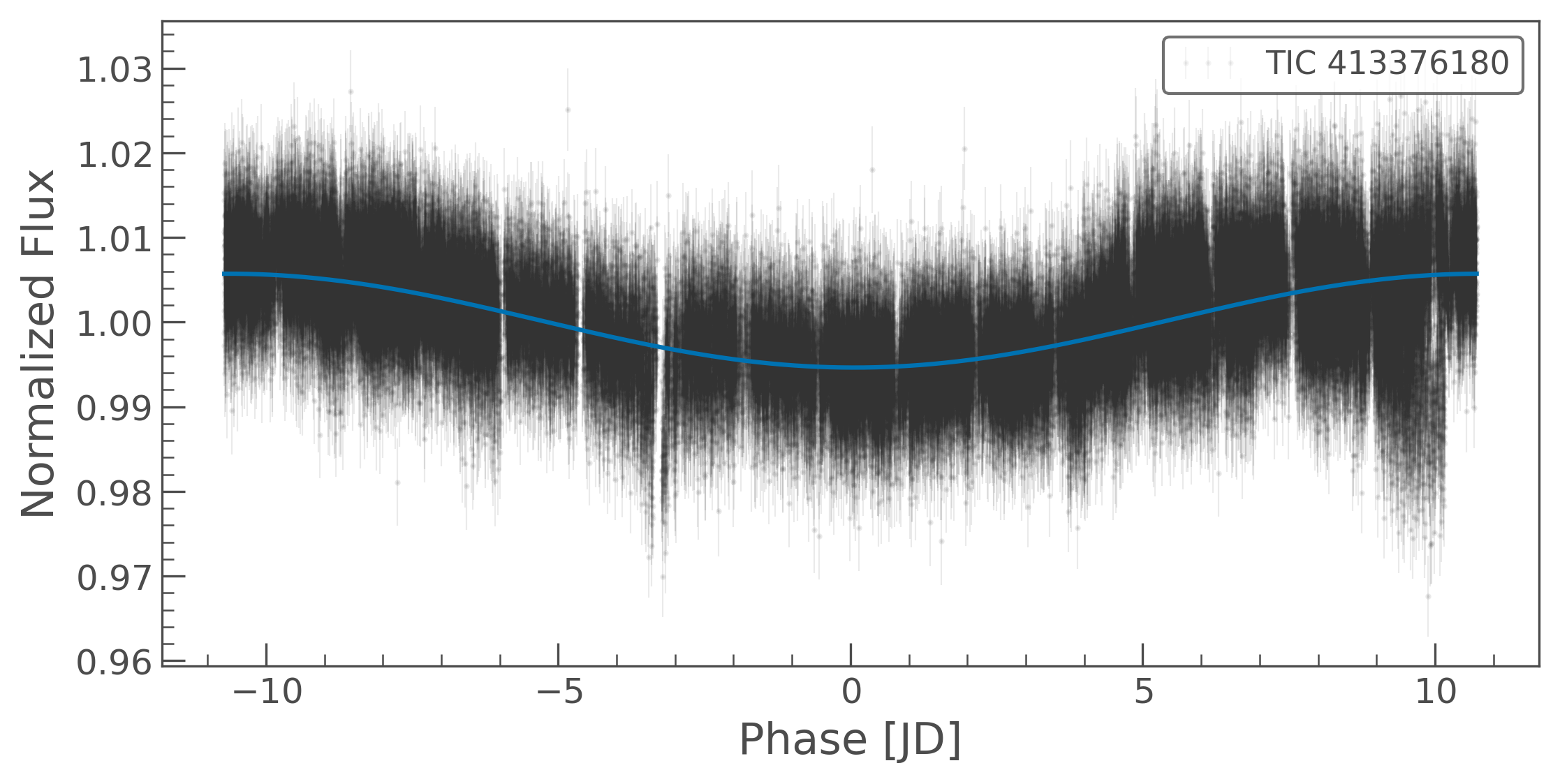}
    \caption{TESS total light curve (Sectors 10, 36, and 63) in phase with its combined GLS peak of 21.4 days (FAP<0.1$\%$ evaluated with the bootstrap method), from the simple aperture photometry (SAP) which was not corrected from long-term trends. The fit of the phased light curve is represented by a blue curve, while the transits of HATS-2 b have been removed.}
    \label{fig:gls}
\end{figure}

\FloatBarrier

\section{PRISM/GEMC and JKTEBOP fits of ExoClock data}

\FloatBarrier

\begin{figure}[h]
    \centering
    \includegraphics[width=\hsize]{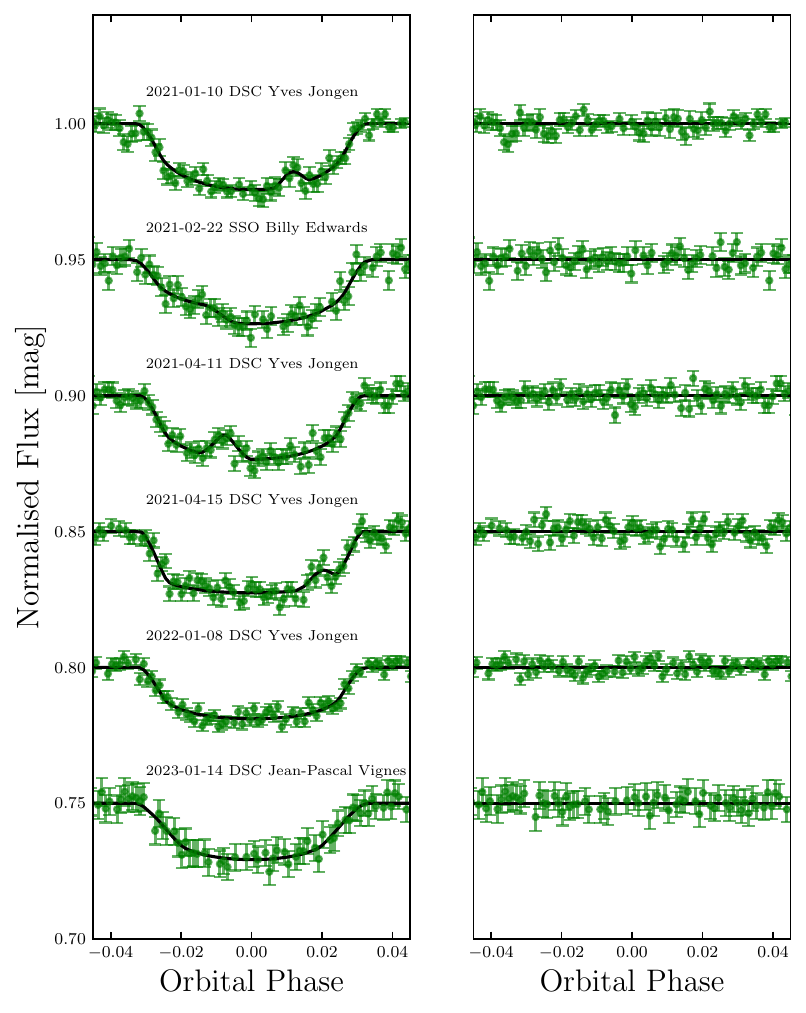}
    \caption{{\it Left}: Light curves of the transits of HATS-2b observed by the astronomers involved in the ExoClock collaboration. They are plotted against the orbital phase and are compared to the best-fitting JKTEBOP and PRISM+GEMC models. The labels indicate in order: observation date, observatory, and observer. DSC stands for Deep Sky Chile, while SSO stands for Sliding Spring Observatory.  {\it Right}: Residuals of each fit.}
    \label{fig:exoclock}
\end{figure}

\begin{table*}
\caption{ Parameters of the PRISM/GEMC and JKTEBOP best fits of the HATS-2 light curves from the ExoClock database.  }
\label{tab:exoclockfit}
\resizebox{\hsize}{!}{  
\centering
\begin{tabular}{c c c c c c c c c c }
\hline \hline \\  [-8pt]
Telescope & Date or ID & Code & $r_{\star} +r_{\rm p}$ & $r_{\rm p} / r_{\star}$ & $i \, (^\circ) $ & $\phi$ ($^{\circ}$) & $\theta$ ($^{\circ}$) & $r_s$ ($^{\circ}$) & $\rho$ \\ 
\hline \\  [-6pt]
DSC & 2021/01/10 & PRISM+GEMC & 0.204 $\pm$ 0.010 &  0.1303 $\pm$ 0.0040 & 88.4 $\pm$ 1.7 & 23.7 $\pm$ 15.0 & 13.0 $\pm$ 20.0 & 5.9 $\pm$ 10.0 & 0.29 $\pm$ 0.15\\
SSO & 2021/02/22 & PRISM+GEMS & 0.216 $\pm$ 0.019 & 0.1268 $\pm$ 0.0059 & 86.4 $\pm$ 2.3 & -60.5 $\pm$ 45.0 & 35.8 $\pm$ 50.0 & 31.2 $\pm$ 20.0 & 0.75 $\pm$ 0.21\\
DSC & 2021/04/11 & PRISM+GEMC & 0.202 $\pm$ 0.013  & 0.1281 $\pm$ 0.0040 & 88.2 $\pm$ 1.9 & -16.8 $\pm$ 2.0 & 11.2 $\pm$ 10.0 & 7.75 $\pm$ 3.57 & 0.46 $\pm$ 0.22\\
DSC & 2021/04/15 & PRISM+GEMC & 0.206 $\pm$ 0.012 & 0.1331 $\pm$ 0.0059 & 87.0 $\pm$ 1.8 & 45.3 $\pm$ 30.0 & 15.8 $\pm$ 30.0 & 8.6 $\pm$ 15.0 & 0.46 $\pm$ 0.29\\
DSC & 2022/01/08 & JKTEBOP & 0.212 $\pm$ 0.016 & 0.1200 $\pm$ 0.0063 & 85.8 $\pm$ 3.4 & -- & -- & -- & --\\
DSC & 2023/01/14 & JKTEBOP & 0.257 $\pm$ 0.028 & 0.1330 $\pm$ 0.0100& 81.5 $\pm$ 2.6 & -- & -- & -- & --\\

\hline
\end{tabular}
 }
\end{table*}

\FloatBarrier

\section{Extra Table}

\FloatBarrier

\begin{table*}
\caption{Parameters of the PRISM/GEMC and JKTEBOP best fits for the HATS-2 light curves used in this work. The final row gives, in bold, the weighted means of the results for the individual data sets.}
\label{tab:fit}
\resizebox{\hsize}{!}{  
\centering
\begin{tabular}{c c c c c c c c c }
\hline \hline \\  [-8pt]
Telescope & Date or ID & Filter & Code & $r_{\star} +r_{\rm p}$ & $r_{\rm p} / r_{\star}$ & $i \, (^\circ) $ & $u_1$ & $u_2$ \\ 
\hline \\  [-6pt]
		MPG 2.2m & 2012/02/28 & $g'$ & PRISM/GEMC & 0.2053 $\pm$ 0.0049 &0.1370 $\pm$ 0.0019 &87.89 $\pm$ 0.93 & 0.82 $\pm$ 0.09 & 0.10 $\pm$ 0.09 \\
		MPG 2.2m & 2012/02/28 & $r'$ & PRISM/GEMC & 0.2059 $\pm$ 0.0060 & 0.1359 $\pm$ 0.0023 & 88.07 $\pm$ 1.11 & 0.69 $\pm$ 0.13  & 0.10 $\pm$ 0.15 \\
		MPG 2.2m & 2012/02/28 & $i'$ & PRISM/GEMC & 0.1990 $\pm$ 0.0044 &0.1342 $\pm$ 0.0017 & 88.48 $\pm$ 1.10 & 0.39 $\pm$ 0.13  & 0.28 $\pm$ 0.23 \\
		MPG 2.2m & 2012/02/28 & $z'$ & PRISM/GEMC & 0.2026 $\pm$ 0.0043 & 0.1370 $\pm$ 0.0023 & 88.46 $\pm$ 1.07 & 0.33 $\pm$ 0.14 & 0.38 $\pm$ 0.25 \\
		MPG 2.2m & 2012/06/01 & $g'$ & PRISM/GEMC &0.1895 $\pm$ 0.0053 &0.1340 $\pm$ 0.0025 & 86.19 $\pm$ 0.82 & 0.41 $\pm$ 0.19 & 0.93 $\pm$ 0.23 \\
		MPG 2.2m & 2012/06/01 & $r'$ & PRISM/GEMC & 0.1927 $\pm$ 0.0038 &0.1329 $\pm$ 0.0016 & 87.41 $\pm$ 0.73 & 0.47 $\pm$ 0.10 & 0.34 $\pm$ 0.18 \\
		MPG 2.2m & 2012/06/01 & $i'$ & PRISM/GEMC & 0.1872 $\pm$ 0.0034 &0.1321 $\pm$ 0.0014 & 88.37 $\pm$ 0.82 & 0.51 $\pm$ 0.09 & 0.07 $\pm$ 0.14 \\
		MPG 2.2m & 2012/06/01 & $z'$ & PRISM/GEMC & 0.1891 $\pm$ 0.0042 &0.1308 $\pm$ 0.0016  & 87.98 $\pm$ 0.89 & 0.32 $\pm$ 0.11 & 0.09 $\pm$ 0.17 \\
		MPG 2.2m & 2015/01/25 & $g'$ & PRISM/GEMC &0.2009 $\pm$ 0.0049  &0.1294 $\pm$ 0.0045 & 87.53 $\pm$ 0.97 & 0.64 $\pm$ 0.10 & 0.06 $\pm$ 0.13 \\
		MPG 2.2m & 2015/01/25 & $r'$ & PRISM/GEMC &0.2080 $\pm$ 0.0052 &0.1300 $\pm$ 0.0026 & 87.42 $\pm$ 0.96 & 0.59 $\pm$ 0.14 & 0.26 $\pm$ 0.20 \\
		MPG 2.2m & 2015/01/25 & $i'$ & PRISM/GEMC &0.2204 $\pm$ 0.0083 &0.1327 $\pm$ 0.0034 & 85.35 $\pm$ 1.17 & 0.33 $\pm$ 0.14 & 0.41 $\pm$ 0.25 \\
		MPG 2.2m & 2015/01/25 & $z'$ & PRISM/GEMC &0.2105 $\pm$ 0.0076 &0.1295 $\pm$ 0.0024 & 87.18 $\pm$ 1.35 & 0.31 $\pm$ 0.13 & 0.57 $\pm$ 0.23 \\
		MPG 2.2m & 2015/01/29 & Combined & PRISM/GEMC &0.1973 $\pm$ 0.0117 &0.1350 $\pm$ 0.0015 & 87.74 $\pm$ 0.99 & 0.50 $\pm$ 0.08 & 0.07 $\pm$ 0.13 \\ [2pt]
		Danish 1.54m & 2016/04/16 & $R$ & PRISM/GEMC &0.2081 $\pm$ 0.0068 &0.1299 $\pm$ 0.0022 & 86.80 $\pm$ 1.10 &0.44 $\pm$ 0.14 &0.45 $\pm$ 0.21 \\
		Danish 1.54m & 2017/05/30 & $R$ & PRISM/GEMC &0.2033 $\pm$ 0.0036 &0.1359 $\pm$ 0.0022 &87.92 $\pm$ 0.96 &0.64 $\pm$ 0.09 &0.10 $\pm$ 0.15 \\
		Danish 1.54m & 2017/07/11 & $R$ & PRISM/GEMC &0.1958 $\pm$ 0.0031 &0.1268 $\pm$ 0.0047 & 89.18 $\pm$ 0.87 &0.57 $\pm$ 0.08 &0.06 $\pm$ 0.14 \\
		Danish 1.54m & 2018/05/17 & $I$ & PRISM/GEMC &0.2004 $\pm$ 0.0054 &0.1312 $\pm$ 0.0021 &88.57 $\pm$ 1.22 &0.53 $\pm$ 0.17 &0.10 $\pm$ 0.24 \\
		Danish 1.54m & 2018/06/05 & $R$ & PRISM/GEMC &0.2000 $\pm$ 0.0059 &0.1297 $\pm$ 0.0021 &88.14 $\pm$ 1.22 &0.47 $\pm$ 0.14 &0.24 $\pm$ 0.22 \\
		Danish 1.54m & 2018/05/24 & $R$ & PRISM/GEMC &0.2024 $\pm$ 0.0052 &0.1324 $\pm$ 0.0020 &88.63 $\pm$ 1.14 &0.75 $\pm$ 0.09 &0.05 $\pm$ 0.11 \\
		Danish 1.54m & 2019/05/19 & $I$ & PRISM/GEMC &0.2066 $\pm$ 0.0100 &0.1257 $\pm$ 0.0038 &87.00 $\pm$ 1.72 &0.22 $\pm$ 0.18 &0.40 $\pm$ 0.24 \\
		Danish 1.54m & 2019/05/23 & $I$ & PRISM/GEMC &0.1968 $\pm$ 0.0031 &0.1321 $\pm$ 0.0024 &88.66 $\pm$ 0.81 &0.45 $\pm$ 0.08 &0.07 $\pm$ 0.11 \\
		Danish 1.54m & 2019/06/07 & $I$ & PRISM/GEMC &0.1996 $\pm$ 0.0036 &0.1311 $\pm$ 0.0021 &88.91 $\pm$ 0.96 &0.44 $\pm$ 0.11 &0.17 $\pm$ 0.20 \\
		Danish 1.54m & 2021/05/26 & $I$ & PRISM/GEMC &0.1994 $\pm$ 0.0047 &0.1336 $\pm$ 0.0026 &88.36 $\pm$ 1.14 &0.31 $\pm$ 0.15 &0.32 $\pm$ 0.25 \\
		Danish 1.54m & 2022/05/13 & $I$ & PRISM/GEMC &0.1889 $\pm$ 0.0208 &0.1301 $\pm$ 0.0028 &88.25 $\pm$ 1.37 &0.16 $\pm$ 0.13 &0.56 $\pm$ 0.24 \\ [2pt]
		TESS s10 & $\#$1 & & JKTEBOP &0.2207 $\pm$ 0.0210 &0.1132 $\pm$ 0.0130 &89.09 $\pm$ 3.34 &0.78 $\pm$ 0.52 &0.38 $\pm$ 0.95 \\
		TESS s10 & $\#$2 & & PRISM/GEMC &0.2097 $\pm$ 0.0289 &0.1322 $\pm$ 0.0066 &87.51 $\pm$ 2.58 &0.21 $\pm$ 0.22 &0.37 $\pm$ 0.23 \\
		TESS s10 & $\#$4 & & PRISM/GEMC &0.2024 $\pm$ 0.0259 &0.1251 $\pm$ 0.0070 &87.21 $\pm$ 2.84 &0.38 $\pm$ 0.22 &0.27 $\pm$ 0.23 \\
		TESS s10 & $\#$5 & & JKTEBOP &0.2378 $\pm$ 0.0492 &0.1320 $\pm$ 0.0156 &83.07 $\pm$ 5.91 &0.61 $\pm$ 0.73 & fixed \\
		TESS s10 & $\#$6 & & JKTEBOP &0.2039 $\pm$ 0.0302 &0.1126 $\pm$ 0.0135 &89.49 $\pm$ 1.67 &0.89 $\pm$ 0.64 &0.13 $\pm$ 1.25 \\
		TESS s10 & $\#$7 & & JKTEBOP &0.2034 $\pm$ 0.0204 &0.1216 $\pm$ 0.0126 &88.89 $\pm$ 3.27 &0.39 $\pm$ 0.85 &0.44 $\pm$ 1.50 \\
		TESS s10 & $\#$8 & & JKTEBOP &0.2159 $\pm$ 0.0315 &0.1186 $\pm$ 0.0051 &85.38 $\pm$ 4.40 & fixed & fixed \\
		TESS s10 & $\#$9 & & JKTEBOP &0.1894 $\pm$ 0.0163 &0.1273 $\pm$ 0.0048 &89.92 $\pm$ 3.11 & fixed & fixed\\
		TESS s10 & $\#$10 & & JKTEBOP &0.2494 $\pm$ 0.0394 &0.1361 $\pm$ 0.0078 &82.08 $\pm$ 3.80 &0.04 $\pm$ 0.45 &fixed \\
		TESS s10 & $\#$11 & & JKTEBOP &0.2418 $\pm$ 0.0494 &0.1398 $\pm$ 0.0117 &82.03 $\pm$ 5.73 &0.24 $\pm$ 0.53 &fixed \\
		TESS s10 & $\#$12 & & JKTEBOP &0.1831 $\pm$ 0.0148 &0.1295 $\pm$ 0.0063 &89.95 $\pm$ 2.54 &0.29 $\pm$ 0.30 &fixed \\
		TESS s10 & $\#$13 & & JKTEBOP &0.1912 $\pm$ 0.0201 &0.1260 $\pm$ 0.0076 &89.59 $\pm$ 3.26 &0.44 $\pm$ 0.40 &fixed \\
		TESS s10 & $\#$14 & & PRISM/GEMC &0.2044 $\pm$ 0.0131 &0.1307 $\pm$ 0.0052 &86.78 $\pm$ 1.80 &0.49 $\pm$ 0.22 &0.25 $\pm$ 0.20 \\
		TESS s10 & $\#$15 & & PRISM/GEMC &0.2100 $\pm$ 0.0247 &0.1290 $\pm$ 0.0066 &87.57 $\pm$ 2.78 &0.71 $\pm$ 0.25 &0.13 $\pm$ 0.23 \\ [2pt]
		TESS s36 & $\#$1 & & JKTEBOP &0.1939 $\pm$ 0.0185 &0.1304 $\pm$ 0.0063 &88.11 $\pm$ 2.91 &0.19 $\pm$ 0.40 &fixed \\
		TESS s36 & $\#$2 & & JKTEBOP &0.2324 $\pm$ 0.0431 &0.1322 $\pm$ 0.0138 &83.55 $\pm$ 5.88 &0.56 $\pm$ 0.87 & fixed \\
		TESS s36 & $\#$3 & & JKTEBOP &0.2604 $\pm$ 0.0447  &0.1373 $\pm$ 0.0071 &81.27 $\pm$ 3.85 & fixed & fixed \\
		TESS s36 & $\#$4 & & JKTEBOP &0.1917 $\pm$ 0.0175 &0.1357 $\pm$ 0.0123 &89.43 $\pm$ 2.50 &0.37 $\pm$ 0.90 &-0.11 $\pm$ 2.45 \\
		TESS s36 & $\#$5 & & JKTEBOP &0.1992 $\pm$ 0.0218 &0.1240 $\pm$ 0.0111 &89.42 $\pm$ 2.68 &0.45 $\pm$ 0.49 &fixed \\
		TESS s36 & $\#$6 & & PRISM/GEMC &0.2057 $\pm$ 0.0170 &0.1289 $\pm$ 0.0067 &87.83 $\pm$ 2.26 &0.20 $\pm$ 0.21 &0.28 $\pm$ 0.24 \\
		TESS s36 & $\#$7 & & PRISM/GEMC &0.2355 $\pm$ 0.0293 &0.1345 $\pm$ 0.0083 &84.69 $\pm$ 2.97 &0.36 $\pm$ 0.23 &0.33 $\pm$ 0.24 \\
		TESS s36 & $\#$8 & & PRISM/GEMC &0.2148 $\pm$ 0.0181 &0.1318 $\pm$ 0.0062 &86.40 $\pm$ 2.32 &0.22 $\pm$ 0.20 &0.36 $\pm$ 0.23 \\
		TESS s36 & $\#$9 & & PRISM/GEMC &0.1948 $\pm$ 0.0139 &0.1291 $\pm$ 0.0063 &88.43 $\pm$ 1.92 &0.24 $\pm$ 0.21 &0.25 $\pm$ 0.23 \\
		TESS s36 & $\#$10 & & PRISM/GEMC &0.2089 $\pm$ 0.0268 &0.1263 $\pm$ 0.0083 &86.99 $\pm$ 2.62 &0.25 $\pm$ 0.23 &0.59 $\pm$ 0.24 \\
		TESS s36 & $\#$11 & & PRISM/GEMC &0.2011 $\pm$ 0.0198 &0.1283 $\pm$ 0.0059 &87.38 $\pm$ 2.44 &0.32 $\pm$ 0.21 &0.37 $\pm$ 0.23 \\
		TESS s36 & $\#$12 & & JKTEBOP &0.1958 $\pm$ 0.0181 &0.1192 $\pm$ 0.0100 &89.94 $\pm$ 2.50 &0.56 $\pm$ 0.43 &fixed \\
		TESS s36 & $\#$13 & & JKTEBOP &0.1938 $\pm$ 0.0149 &0.1350 $\pm$ 0.0030 &89.17 $\pm$ 2.86 &-0.09 $\pm$ 0.27 &fixed \\
		TESS s36 & $\#$14 & & JKTEBOP &0.1942 $\pm$ 0.0230 &0.1243 $\pm$ 0.0133 &89.82 $\pm$ 2.93 &0.41 $\pm$ 0.83 &0.43 $\pm$ 1.80 \\
		TESS s36 & $\#$15 & & JKTEBOP &0.1968 $\pm$ 0.0170 &0.1249 $\pm$ 0.0082 &89.87 $\pm$ 2.38 &0.46 $\pm$ 0.37 &fixed \\
        TESS s63 & $\#$1 & & PRISM/GEMC &0.2017 $\pm$ 0.0172 &0.1301 $\pm$ 0.0044 &88.47 $\pm$ 2.24 &0.32 $\pm$ 0.20 & 0.21 $\pm$ 0.23 \\
        TESS s63 & $\#$2 & & PRISM/GEMC &0.2082 $\pm$ 0.0234 &0.1320 $\pm$ 0.0047 &86.21 $\pm$ 2.64 &0.19 $\pm$ 0.20 & 0.19 $\pm$ 0.23 \\
        TESS s63 & $\#$3 & & PRISM/GEMC &0.1967 $\pm$ 0.0167 &0.1376 $\pm$ 0.0067 &87.65 $\pm$ 2.14 &0.16 $\pm$ 0.14 & 0.27 $\pm$ 0.23 \\
        TESS s63 & $\#$4 & & PRISM/GEMC &0.2214 $\pm$ 0.0512 &0.1190 $\pm$ 0.0143 &85.88 $\pm$ 4.63 &0.76 $\pm$ 0.26 & 0.11 $\pm$ 0.22 \\
        TESS s63 & $\#$5 & & PRISM/GEMC &0.2220 $\pm$ 0.0459 &0.1195 $\pm$ 0.0139 &85.74 $\pm$ 4.30 &0.74 $\pm$ 0.25 & 0.13 $\pm$ 0.23 \\
        TESS s63 & $\#$6 & & PRISM/GEMC &0.2230 $\pm$ 0.0264 &0.1297 $\pm$ 0.0053 &84.53 $\pm$ 2.95 &0.23 $\pm$ 0.21 &f 0.34 $\pm$ 0.24 \\
        TESS s63 & $\#$7 & & PRISM/GEMC &0.1913 $\pm$ 0.0096 &0.1381 $\pm$ 0.0042 &88.80 $\pm$ 1.60 &0.14 $\pm$ 0.16 & 0.14 $\pm$ 0.22 \\
        TESS s63 & $\#$8 & & PRISM/GEMC &0.2538 $\pm$ 0.0302 &0.1393 $\pm$ 0.0061 &81.29 $\pm$ 3.21 &0.10 $\pm$ 0.20 & 0.18 $\pm$ 0.24 \\
        TESS s63 & $\#$9 & & PRISM/GEMC &0.2035 $\pm$ 0.0157 &0.1319 $\pm$ 0.0064 &88.12 $\pm$ 2.17 &0.22 $\pm$ 0.20 & 0.38 $\pm$ 0.24 \\
        TESS s63 & $\#$10 & & PRISM/GEMC &0.2302 $\pm$ 0.0266 &0.1338 $\pm$ 0.0051 &84.21 $\pm$ 2.91 &0.09 $\pm$ 0.18 & 0.20 $\pm$ 0.24 \\
        TESS s63 & $\#$11 & & PRISM/GEMC &0.2022 $\pm$ 0.0271 &0.1243 $\pm$ 0.0066 &87.32 $\pm$ 2.91 &0.41 $\pm$ 0.21 & 0.31 $\pm$ 0.24 \\
        TESS s63 & $\#$12 & & PRISM/GEMC &0.2016 $\pm$ 0.0244 &0.1297 $\pm$ 0.0050 &88.26 $\pm$ 2.75 &0.31 $\pm$ 0.21 & 0.31 $\pm$ 0.24 \\
        TESS s63 & $\#$13 & & PRISM/GEMC &0.2065 $\pm$ 0.0222 &0.1351 $\pm$ 0.0061 &86.87 $\pm$ 2.45 &0.25 $\pm$ 0.18 & 0.25 $\pm$ 0.24\\
        TESS s63 & $\#$14 & & PRISM/GEMC &0.2128 $\pm$ 0.0220 &0.1330 $\pm$ 0.0057 &86.00 $\pm$ 2.54 &0.21 $\pm$ 0.20 & 0.23 $\pm$ 0.23 \\
        TESS s63 & $\#$15 & & PRISM/GEMC &0.2076 $\pm$ 0.0282 &0.1283 $\pm$ 0.0050 &88.37 $\pm$ 2.98 &0.27 $\pm$ 0.21 & 0.17 $\pm$ 0.23 \\
        TESS s63 & $\#$16 & & PRISM/GEMC &0.1937 $\pm$ 0.0270 &0.1314 $\pm$ 0.0057 &87.59 $\pm$ 2.82 &0.49 $\pm$ 0.23 & 0.21 $\pm$ 0.24 \\
        TESS s63 & $\#$17 & & PRISM/GEMC &0.1986 $\pm$ 0.0293 &0.1302 $\pm$ 0.0059 &87.60 $\pm$ 3.09 &0.39 $\pm$ 0.22 & 0.35 $\pm$ 0.23 \\
        TESS s63 & $\#$18 & & PRISM/GEMC &0.2054 $\pm$ 0.0380 &0.1311 $\pm$ 0.0081 &88.55 $\pm$ 3.74 &0.59 $\pm$ 0.24 & 0.16 $\pm$ 0.23 \\ [2pt]
		TESS s10 & combined $\#1-\#8$ & & JKTEBOP & 0.2020 $\pm$ 0.0121 & 0.1272 $\pm$ 0.0031 & 87.60 $\pm$ 2.61 & 0.37 $\pm$ 0.13 & fixed \\
		TESS s10 & ~~combined $\#9-\#15$ & & JKTEBOP & 0.2115 $\pm$ 0.0144 & 0.1346 $\pm$ 0.0029 & 85.47 $\pm$ 2.81 & 0.24 $\pm$ 0.13 & fixed \\
		TESS s36 & combined $\#1-\#8$ & & JKTEBOP & 0.2021 $\pm$ 0.0089 & 0.1318 $\pm$ 0.0029 & 87.73 $\pm$ 2.21 & 0.26 $\pm$ 0.14 & fixed \\
		TESS s36 & ~~combined $\#9-\#15$ & & JKTEBOP & 0.1959 $\pm$ 0.0091 & 0.1296 $\pm$ 0.0024 & 89.24 $\pm$ 2.15 & 0.28 $\pm$ 0.12 & fixed \\
        TESS s63 & combined $\#$1-$\#$5 & & JKTEBOP & 0.2050 $\pm$ 0.0170 & 0.1246 $\pm$ 0.0030 & 86.6 $\pm$ 3.2 & 0.28 $\pm$ 0.18 & fixed \\
        TESS s63& combined $\#$6-$\#$9 && JKTEBOP & 0.1990 $\pm$ 0.0190 & 0.1271 $\pm$ 0.0027 & 87.0 $\pm$ 3.2 & 0.14 $\pm$ 0.20 & fixed \\
        TESS s63& combined $\#$10-$\#$14 & & JKTEBOP & 0.2040 $\pm$ 0.0160  & 0.1260 $\pm$ 0.0025 & 86.7 $\pm$ 3.0 & 0.18 $\pm$ 0.20 & fixed \\
        TESS s63 & combined $\#$15-$\#$18 & & JKTEBOP & 0.1970 $\pm$ 0.0110 & 0.1210 $\pm$ 0.0037 & 89.8 $\pm$ 1.8 & 0.54 $\pm$ 0.14 & fixed \\
		\hline \\[-6pt]
\textbf{Final Results} & & & & \textbf{0.19901 $\pm$ 0.00090} & \textbf{0.13164 $\pm$ 0.00039} & \textbf{87.83 $\pm$ 0.18   } & & \\[2pt]
\hline
\end{tabular}
 }
\end{table*}

\FloatBarrier

\end{appendix}

\end{document}